\documentclass{aa}  

\usepackage{graphicx}
\usepackage{txfonts}
\usepackage{hyperref}
\usepackage{color}
\usepackage{float}

\begin{document} 

\title{Magnetised Polish doughnuts revisited}

\author{Sergio Gimeno-Soler\inst{1}   \and Jos\'e A.~Font\inst{1,2} }
\institute{Departamento de Astronom\'{\i}a y Astrof\'{\i}sica, Universitat de Val\`encia, Dr. Moliner 50, 46100, Burjassot (Val\`encia), Spain.\\\email{sergio.gimeno@uv.es}
\and
Observatori Astron\`omic, Universitat de Val\`encia, C/ Catedr\'atico Jos\'e Beltr\'an 2, 46980, Paterna (Val\`encia), Spain. \\
\email{j.antonio.font@uv.es}
}

   \date{}
 
  \abstract
{We discuss a procedure to build new sequences of magnetised, equilibrium tori around Kerr black holes which combines two approaches previously considered in the literature. For simplicity we assume that the test-fluid approximation holds, and hence we neglect the self-gravity of the fluid. The models are built assuming a particular form of the angular momentum distribution from which the location and morphology of equipotential surfaces can be computed. This ansatz includes, in particular, the  constant angular momentum case originally employed in the construction of thick tori -- or Polish doughnuts -- and it has already been used to build equilibrium sequences of purely hydrodynamical models. We discuss the properties of the new models and their dependence on the initial parameters. These new sequences can be used as initial data for magnetohydrodynamical evolutions in general relativity.}


   \maketitle

\section{Introduction}

Matter accretion on to black holes is the most efficient form of energy production known in nature. The conversion of the
gravitational energy of the infalling matter into heat and radiation may reach efficiencies of about 43\% in the case of maximally rotating (Kerr) black holes. For this reason, systems formed by a black hole surrounded by an accretion disc are deemed responsible for many of the most energetic astronomical phenomena observed in the cosmos. In particular, (geometrically) thick accretion discs (or tori) are believed to be present in quasars and other active galactic nuclei, some X-ray binaries and microquasars, as well as in the central engine of gamma-ray bursts.
The latter are alluded to in connection with mergers of neutron star binaries and black hole neutron star binaries, as well as with the rotational collapse that ensues at the end of the life of some massive stars. As numerical simulations show, such events  often result in a black hole surrounded by a torus (see e.g.~\cite{Rezzolla:2010,Sekiguchi:2011,Faber:2012,Shibata:2011,Baiotti:2017}).

The investigation of this type of systems, either by analytical or numerical means, may rely on the ability to construct suitable 
representations based on physical assumptions. The construction of equilibrium models of stationary discs around black holes has indeed a long tradition (see~\cite{Abramowicz:2013} and references therein). In particular, the so-called ``Polish doughnuts"~\citep{Abramowicz:1978,Kozlowski:1978} provide a  very general method to build equilibrium configurations of perfect fluid matter orbiting around a Kerr black hole. This fully relativistic model assumes that the disc is non-magnetised and that the matter obeys a  constant specific angular momentum distribution. This method was later extended by~\cite{Komissarov:2006}  by adding a purely azimuthal magnetic field to build magnetised tori around rotating black holes. Dynamical evolutions of magnetized tori built with the Komissarov solution were first reported by~\cite{Montero:2007}.
On the other hand, assuming different distributions of angular momentum in the discs, \citet{Qian:2009} presented a method to build sequences of black hole thick accretion discs in dynamical equilibria, restricted however to the purely hydrodynamical case. 
 
 In this paper we combine the two approaches considered in~\cite{Komissarov:2006} and~\citet{Qian:2009} to build new sequences of equilibrium tori around Kerr black holes. Building on these works, we present here the extension of the models of~\citet{Qian:2009} to account for discs endowed with a purely toroidal magnetic field. In our procedure we hence assume a form of the angular momentum distribution that departs from the constant case considered by~\cite{Komissarov:2006} and from which the location and morphology of the equipotential surfaces can be numerically computed. As we shall show below, for the particular case of constant angular momentum distributions, our method is in good agreement with the results of~\citet{Komissarov:2006}. We also note the recent work of~\citet{Wielgus:2015} where Komissarov's solution was extended for the particular case of power-law distributions of angular momentum. Moreover, the magnetised tori of~\citet{Wielgus:2015} were used to explore the growth of the magneto-rotational instabilty (MRI) through time-dependent numerical simulations. In particular, the long-term evolution of those tori has been recently investigated by~\citet{Fragile:2017}, who paid special attention to the decay of their magnetisation. 

The organization of the paper is as follows: Section~\ref{framework} presents the analytic framework to build the discs while Section~\ref{methodology} explains the corresponding numerical procedure. The sequences of models are discussed in Section~\ref{results}. Finally, the conclusions are summarized in Section~\ref{conclusions}, where we also briefly indicate potential extensions of this work we plan to inspect. In our work we assume that the test-fluid approximation holds, thus neglecting the self-gravity of the fluid, and further assume that the spacetime is described by the Kerr metric. In mathematical expressions below Greek indices are spacetime indices running from 0 to 3 and Latin indices are only spatial. We use geometrized units where $G = c = 1$.

\section{Framework}
\label{framework}

Equilibrium tori around Kerr black holes are built assuming that the spacetime gravitational potentials and the fluid fields are stationary and axisymmetric. In all the derivations presented below, the Kerr metric is implicitly written using standard Boyer-Lindquist coordinates. It is convenient to introduce a number of relevant characteristic radii, such as the radius
of the marginally stable circular orbit, $r_{\rm ms}$, and the radius of the marginally bound circular orbit, $r_{\rm mb}$, given by
\begin{eqnarray}
r_{\rm ms} &=& M\,\left(3+Z_2-\left[(3-Z_1)(3+Z_1+2Z_2)\right]^{1/2})\right)\,,
\\
r_{\rm mb} &=& 2M\,\left(  1-\frac{a_*}{2} + \sqrt{1-a_*} \right)\,,
\end{eqnarray}
where we have defined the following quantities, $Z_1=1+(1-a_*^2)^{1/3}[(1+a_*)^{1/3}+(1-a_*)^{1/3}]$, $Z_2=(3a_*^2+Z_1^2)^{1/2}$, and $a_*=a/M$, with $a$ and $M$ being the spin Kerr parameter and the black hole mass, respectively.

\subsection{Distribution of angular momentum}

We introduce the specific angular momentum $l$ and the angular velocity $\Omega$ employing the standard definitions,
\begin{equation}
l = - \frac{u_{\phi}}{u_t}, \;\;\; \Omega = \frac{u^{\phi}}{u^t},
\end{equation}
where $u^{\mu}$ is the fluid four-velocity.
The relationship between $l$ and $\Omega$ is given by the equations
\begin{equation}
l = - \frac{\Omega g_{\phi\phi} + g_{t\phi}}{\Omega g_{t\phi} + g_{tt}}, \;\;\; \Omega = - \frac{l g_{tt} + g_{t\phi}}{l g_{t\phi} + g_{\phi\phi}},
\end{equation}
where $g_{\mu\nu}$ is the metric tensor and we are assuming circular motion, i.e. the four-velocity can be written as
\begin{equation}
u^{\mu} = (u^t, 0, 0, u^{\phi})\,.
\end{equation}
We also introduce the Keplerian angular momentum (for prograde motion) in the equatorial plane $l_{\mathrm{K}}$, defined as 
\begin{equation}\label{eq:kepler}
l_{\mathrm{K}}(r) = \frac{M^{1/2}(r^{2}-2aM^{1/2}r^{1/2}+a^{2})}{r^{3/2}-2Mr^{1/2}+aM^{1/2}}\,.
\end{equation}

\cite{Jaroszynski:1980} argued that the slope of the specific angular momentum should range between two limiting cases, namely $l = \mathrm{const.}$ and $\Omega = \mathrm{const}$. Following \citet{Qian:2009} we assume an angular momentum distribution ansatz given by  
\begin{equation}
l (r,\theta) = \left\{ \label{eq:ansatz} 
  \begin{array}{lr}
    l_0 \left(\frac{l_{\mathrm{K}}(r)}{l_0}\right)^{\beta}\sin^{2\gamma}{\theta} &  \text{for } r \geq r_{\mathrm{ms}}\\
    l_{\mathrm{ms}}(r)\sin^{2\gamma}{\theta} & \text{for } r < r_{\mathrm{ms}}
  \end{array}
\right.
\end{equation}
where constants $l_0$ and $l_{\mathrm{ms}}(r)$ are defined by $l_0 \equiv \eta l_{\mathrm{K}}(r_{\mathrm{ms}})$ and $l_{\mathrm{ms}}(r) \equiv l_0 [l_{\mathrm{K}}(r_{\mathrm{ms}})/l_0]^{\beta}$. Therefore, the model for the distribution of angular momentum has three free parameters, $\beta$ , $\gamma$ and $\eta$, whose range of variation is given by~\citep{Qian:2009}
\begin{equation}
0 \leq \beta \leq 1, \quad -1 \leq \gamma \leq 1, \quad 1 \leq \eta \leq \eta_{\mathrm{max}},
\end{equation}
with $\eta_{\mathrm{max}} = l_{\mathrm{K}}(r_{\mathrm{mb}})/l_{\mathrm{K}}(r_{\mathrm{ms}})$. In this paper, and as it is done for hydrodynamical discs in~\citet{Qian:2009}, we choose $\eta = \eta_{\mathrm{max}}$, and then we can write $l_0$ as $l_0 = l_{\mathrm{K}}(r_{\mathrm{mb}})$. For this choice of $\eta$, we can find the location of the cusp of the disc  within the range $r_{\mathrm{mb}} \leq r_{\mathrm{cusp}} \leq r_{\mathrm{ms}}$ (for $0 \leq \beta \leq 1$). The cusp is defined as the circle in the equatorial plane on which the pressure gradient vanishes and the angular momentum of the disc equals the Keplerian angular momentum. Moreover, this value of $\eta$ guarantees that constant angular momentum discs ($\beta = \gamma = 0$) with their inner edge located at the cusp ($r_{\mathrm{in}} = r_{\mathrm{cusp}}$) have no outer boundary, i.e~they are infinite discs (see also~\cite{Font:2002}). 

\subsection{Magnetised discs}

The equations of ideal general relativistic MHD are the following conservation laws, $\nabla_{\mu} T^{\mu\nu} = 0$, $\nabla_{\mu} \,^\ast F^{\mu\nu} = 0$, and 
$\nabla_{\mu} (\rho u^{\mu}) = 0$, 
where $\nabla_{\mu}$ is the covariant derivative and
\begin{equation}\label{eq:e-m_tensor}
T^{\mu\nu} = (w + b^2)u^{\mu}u^{\nu} + \left(p + \frac{1}{2}b^2\right)g^{\mu\nu} - b^{\mu}b^{\nu},
\end{equation}
is the energy-momentum tensor of a magnetised perfect fluid, $w$ and $p$ being the fluid enthalpy density and fluid pressure, respectively. 
Moreover, $^\ast F^{\mu\nu} = b^{\mu}u^{\nu} - b^{\nu}u^{\mu}$ is the (dual of the) Faraday tensor relative to an observer with 
four-velocity $u^{\mu}$, and $b^{\mu}$ is the magnetic field in that frame, with
$b^2=b^{\mu}b_{\mu}$. Assuming the magnetic field is purely azimuthal, i.e.~$b^r = b^{\theta} = 0$,
and taking into account that the flow is stationary and axisymmetric, the conservation of the current density and of the Faraday tensor follow. Contracting Eq.~\eqref{eq:e-m_tensor} with the projection tensor $h^{\alpha}_{\,\,\beta} = \delta^{\alpha}_{\,\,\beta} + u^{\alpha}u_{\beta}$, we arrive at
\begin{equation}
(w + b^2)u_{\nu}\partial_i u^{\nu} + \partial_i\left(p + \frac{b^2}{2}\right) - b_{\nu}\partial_i b^{\nu}=0\,,
\end{equation}
where $i = r, \theta$. Following~\cite{Komissarov:2006} we rewrite this equation in terms of the specific angular momentum $l$ and of the angular velocity $\Omega$, to obtain
\begin{equation}\label{eq:diff_ver}
\partial_i(\ln u_t|) - \frac{\Omega \partial_i l}{1-l\Omega} + \frac{\partial_i p}{w} + \frac{\partial_i(\mathcal{L}b^2)}{2\mathcal{L}w} = 0\,,
\end{equation}
where $\mathcal{L} = g_{t\phi}^2 - g_{tt}g_{\phi\phi}$.
To integrate Eq.~\eqref{eq:diff_ver} we first assume a barotropic equation of state $w = w(p)$ of the form
\begin{equation}\label{eq:eos_fluid}
p = K w^{\kappa},
\end{equation}
with $K$ and $\kappa$ constants.
Then, we define the magnetic pressure as $p_{\mathrm{m}} = b^2/2$, and introduce the definitions $\tilde{p}_{\mathrm{m}} = \mathcal{L} p_{\mathrm{m}}$ and $\tilde{w} = \mathcal{L} w$, in order to write an analogue equation to Eq.~\eqref{eq:eos_fluid} for $\tilde{p}_{\mathrm{m}}$~\citep{Komissarov:2006}
\begin{equation}\label{eq:eos_mag_tilde}
\tilde{p}_{\mathrm{m}} = K_{\mathrm{m}} \tilde{w}_{\mathrm{m}}^{\lambda
},
\end{equation}
or, in terms of the magnetic pressure $p_{\mathrm{m}}$
\begin{equation}\label{eq:eos_mag}
p_{\mathrm{m}} = K_{\mathrm{m}} \mathcal{L}^{\lambda
-1} w^{\lambda
},
\end{equation}
where $K_{\mathrm{m}}$ and $\lambda
$ are constants.
This particular choice of barotropic relationships, $w = w(p)$ and $\tilde{w} = \tilde{w}(\tilde{p}_{\mathrm{m}})$, fulfill the general relativistic version of the von Zeipel theorem for a toroidal magnetic field~\citep{vonZeipel:1924, Zanotti:2015}, i.e.~the surfaces of constant $\Omega$ and constant $l$ coincide.

We can now integrate Eq.~\eqref{eq:diff_ver} to obtain
\begin{equation}\label{eq:pre_full_int}
\ln |u_t| - \int^l_0 \frac{\Omega \mathrm{d}l}{1 - \Omega l} + \int^p_0 \frac{\mathrm{d}p}{w} + \int_0^{\tilde{p}_{\mathrm{m}}} \frac{\mathrm{d}\tilde{p}_{\mathrm{m}}}{\tilde{w}} = \mathrm{const}.
\end{equation}
On the surface of the disc, and particularly on its inner edge, the conditions
$p = \tilde{p}_{\mathrm{m}} = 0, \; u_t = u_{t, \mathrm{in}}, \; l = l_{\mathrm{in}}$
are satisfied and, therefore, the integration constant is simply given by
\begin{equation}
\mathrm{const.} = \ln |u_t| - \int^l_{l_\mathrm{in}} \frac{\Omega \mathrm{d}l}{1 - \Omega l}\,.
\end{equation}
We can also introduce the total (gravitational plus centrifugal) potential $W$~\citep{Abramowicz:1978} and write the integral form of the equation of motion (the relativistic Euler equation) as
\begin{equation}\label{eq:potential}
W - W_{\mathrm{in}} = \ln|u_t| - \ln|u_{t,\mathrm{in}}| - \int^{l}_{l_{\mathrm{in}}} \frac{\Omega \mathrm{d}l}{1 - \Omega l},
\end{equation}
where $W_{\mathrm{in}}$ is the potential at the inner edge of the disc (in the equatorial plane). With this definition, we can write Eq.~\eqref{eq:pre_full_int} as
\begin{equation}\label{eq:full_int}
W - W_{\mathrm{in}} = \int^p_0 \frac{\mathrm{d}p}{w} + \int_0^{\tilde{p}_{\mathrm{m}}} \frac{\mathrm{d}\tilde{p}_{\mathrm{m}}}{\tilde{w}},
\end{equation}
which for a barotropic equation of state can be easily integrated to give
\begin{equation}\label{eq:pre_enthalpy_eq}
W - W_{\rm{in}} + \frac{\kappa}{\kappa - 1}\frac{p}{w} + \frac{\lambda
}{\lambda
 - 1}\frac{p_{\mathrm{m}}}{w} = 0\,.
\end{equation}
Replacing $p$ and $p_{\mathrm{m}}$ by equations \eqref{eq:eos_fluid} and \eqref{eq:eos_mag}, the previous equation reduces to
\begin{equation}\label{eq:enthalpy_eq}
W - W_{\rm{in}} + \frac{\kappa}{\kappa - 1} K w^{\kappa - 1} + \frac{\lambda
}{\lambda
 - 1}K_{\mathrm{m}}(\mathcal{L} w)^{\lambda
 - 1} = 0,
\end{equation}
which relates the distribution of the potential with the distribution of the enthalpy density.

\section{Methodology}
\label{methodology}

To construct our models of magnetised discs we follow the approach described in \citet{Qian:2009}. First, we find the radial location of the cusp and of the centre of the disc in the equatorial plane, $r_{\mathrm{cusp}}$ and $r_{\mathrm{c}}$, defined as the solutions to the equation $l(r) - l_{\mathrm{K}} = 0$.
Next, we compute the partial derivatives of the potential, Eq.~\eqref{eq:potential}
\begin{equation}\label{eq:radial_der_pot}
\partial_r W = \partial_r \ln|u_t| - \frac{\Omega \partial_rl}{1 - \Omega l}\,,
\end{equation}
and
\begin{equation}\label{eq:polar_der_pot}
\partial_{\theta} W = \partial_{\theta} \ln|u_t| - \frac{\Omega \partial_{\theta}l}{1 - \Omega l}\,.
\end{equation}
Then, we integrate the radial partial derivative of the potential along the segment $[r_{\mathrm{cusp}}, r_{\mathrm{c}}]$ (assuming $W_{\mathrm{cusp}} = 0$) at the equatorial plane, thus obtaining the equatorial distribution of the potential between $r_{\mathrm{cusp}}$ and $r_{\mathrm{c}}$
\begin{equation}\label{eq:equatorial_pot}
W_{\mathrm{eq}}(r) = \int^{r_{\mathrm{c}}}_{r_{\mathrm{cusp}}}\left(\partial_r \ln|u_t| - \frac{\Omega \partial_rl}{1 - \Omega l}\right).
\end{equation}
Following~\citet{Qian:2009} (see also~\citet{Jaroszynski:1980}) we can divide equations \eqref{eq:radial_der_pot} and \eqref{eq:polar_der_pot} to obtain
\begin{equation}\label{eq:F}
F(r, \theta) = -\frac{\partial_r W}{\partial_{\theta} W} = \frac{\mathrm{d}\theta}{\mathrm{d}r}\,,
\end{equation}
where function $F(r, \theta)$ is known in closed form for the Kerr metric once an assumption for the angular momentum distribution has been made (cf.~Eq.~(\ref{eq:ansatz})). For this reason, Eq.~(\ref{eq:F}) also takes the form of an ordinary differential equation for the surfaces of constant potential, $\theta=\theta(r)$, which, upon integration, yields the location of those surfaces. Note that in~\citet{Qian:2009} these are surfaces of constant {\it fluid pressure} instead, since their discs are purely hydrodynamical.

\begin{figure*}[t]
\centering
\includegraphics[scale=0.14]{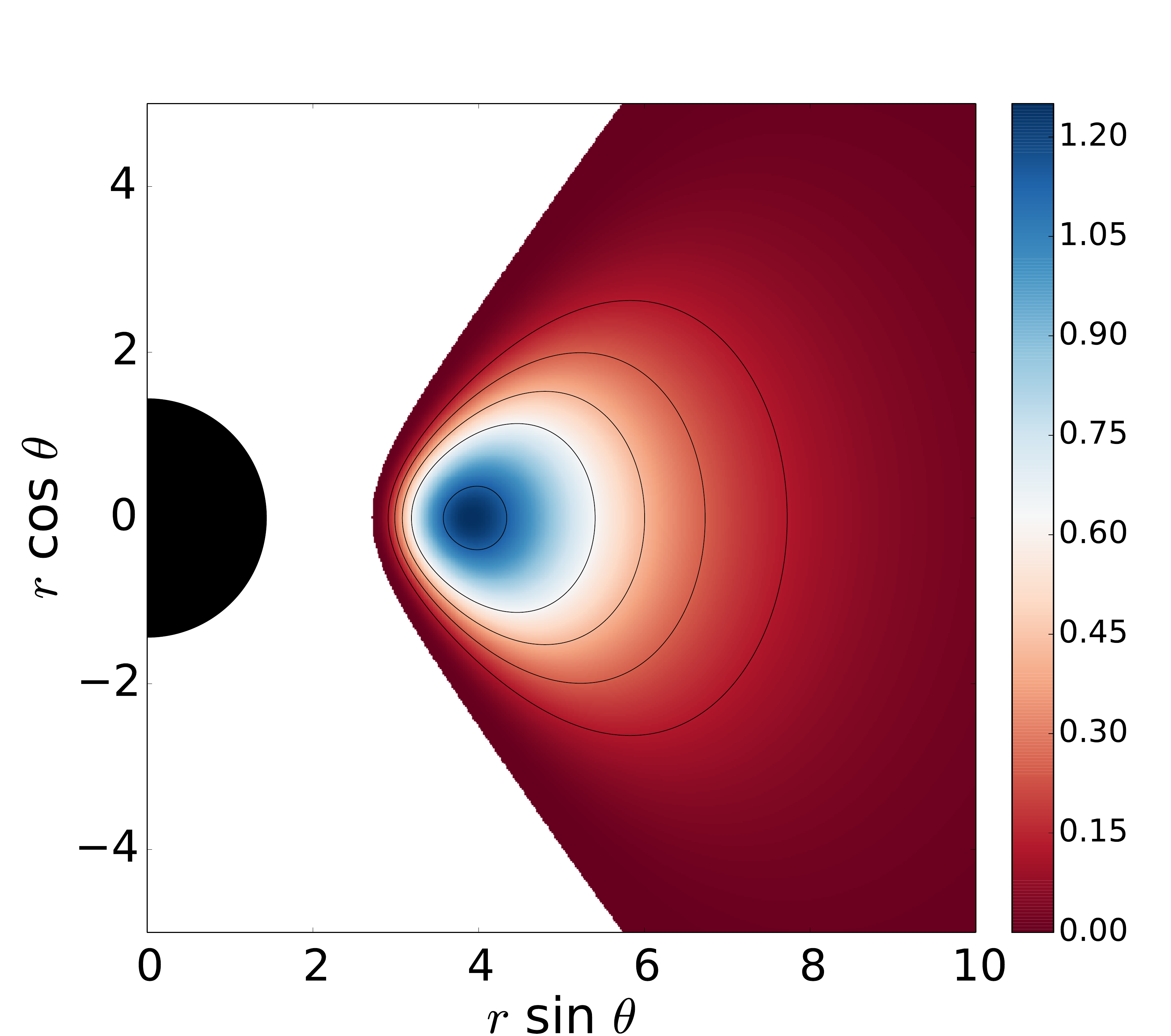}
\hspace{-0.3cm}
\includegraphics[scale=0.14]{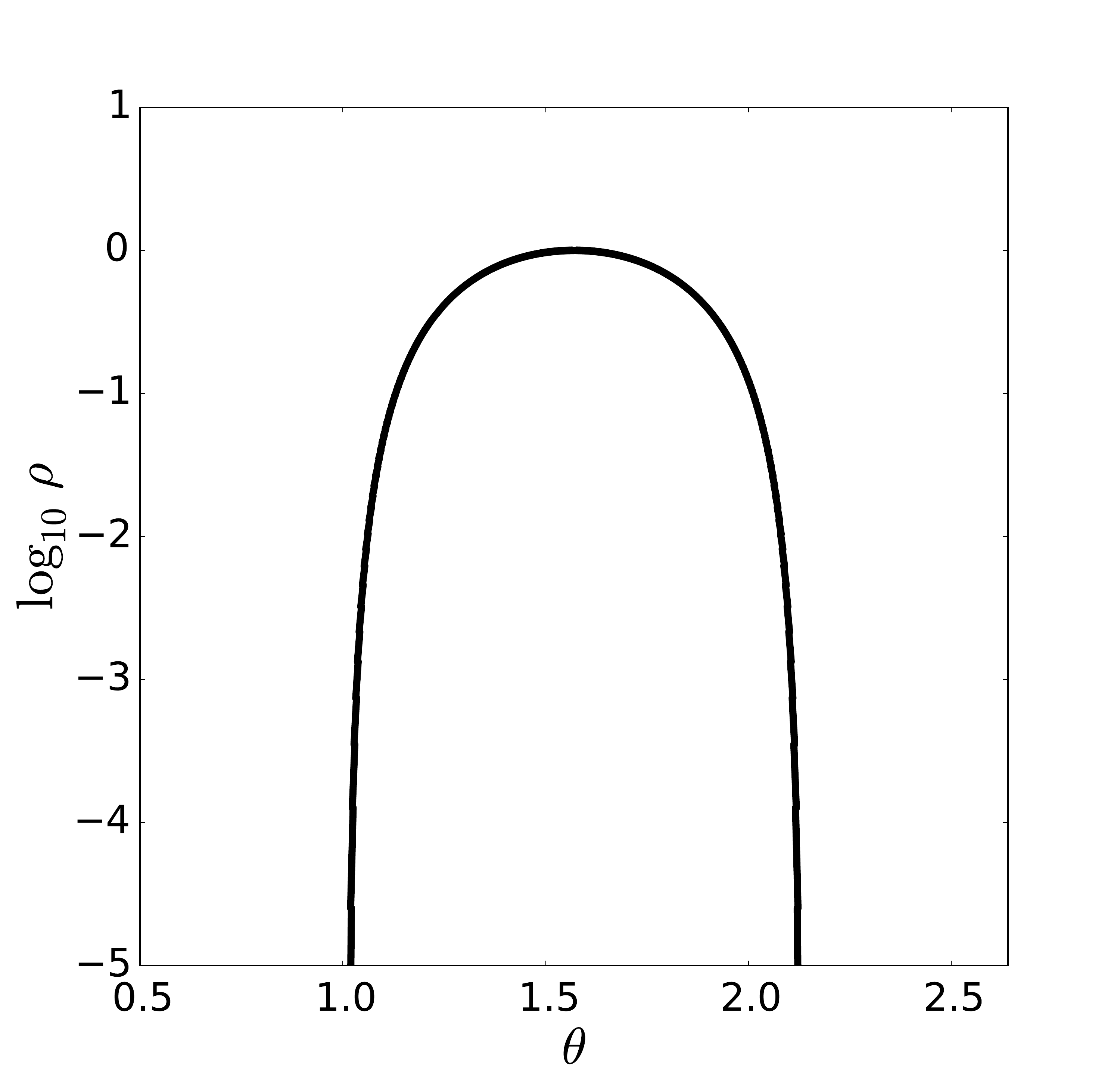}
\hspace{-0.2cm}
\includegraphics[scale=0.14]{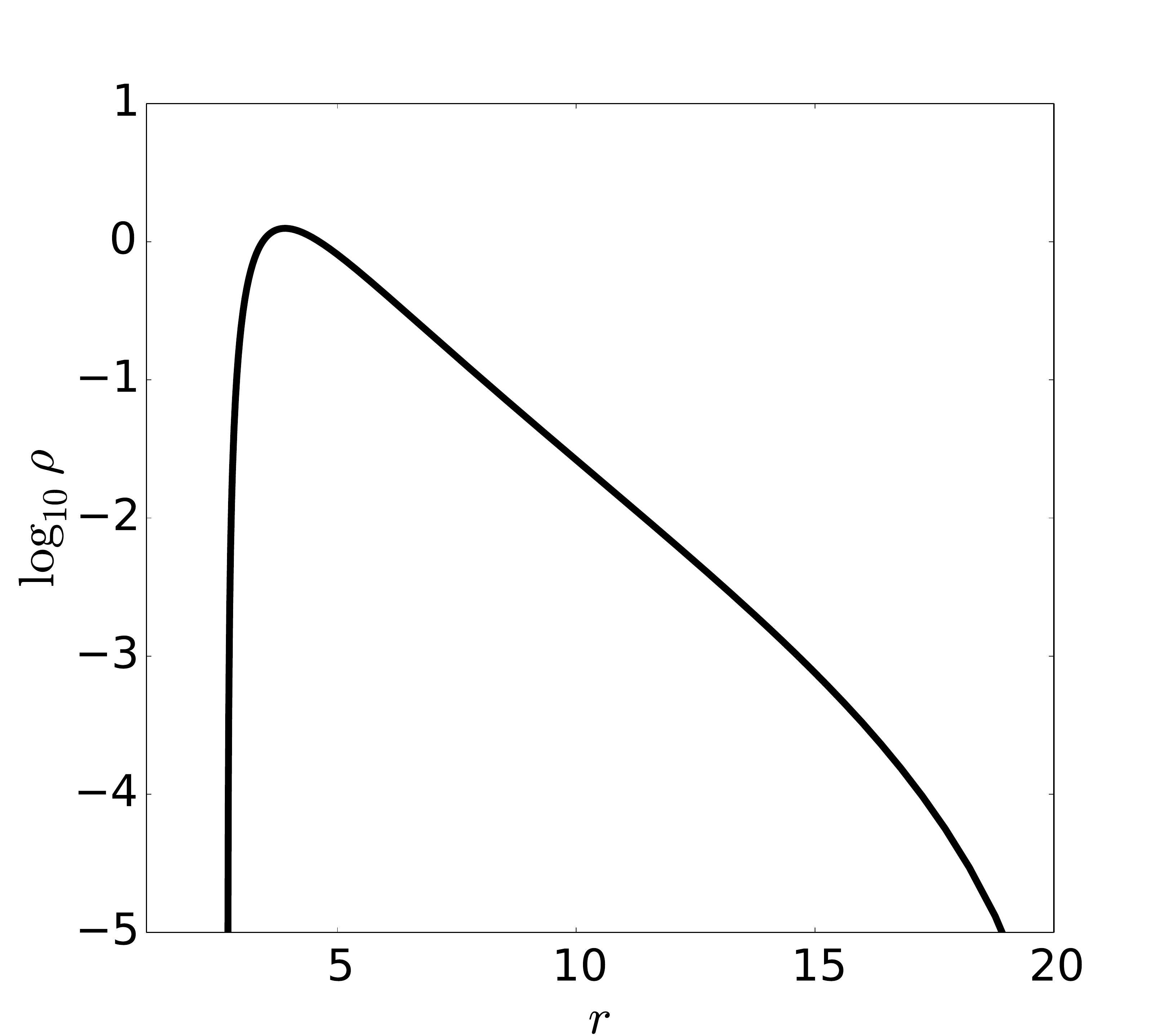}
\caption{Comparison with Komissarov's solution for a constant angular momentum model. The left panel shows the rest-mass density distribution in logarithm scale for all radii and all angles while the middle and right panels show, respectively, the angular profile of the logarithm of the rest-mass density at the centre of the disc and the radial profile of the logarithm of the rest-mass density at the equatorial plane.}
\label{komissarov}
\end{figure*}

Next, we choose all the initial radial values for the integration of Eq.~\eqref{eq:F} to lie between $r_{\mathrm{cusp}}$ and $r_{\mathrm{c}}$ ($\theta = \pi / 2$). Since we are only interested in the equipotential surfaces inside the Roche lobe of the disc, our choice of initial values provides us a mapping of the equipotential surfaces of the torus. Given that we have already obtained both the equipotential surfaces $\theta(r)$ which cross the segment $[r_{\mathrm{cusp}}, r_{\mathrm{c}}]$ at the equatorial plane and the values of the potential in that segment, we can obtain the complete potential distribution for the torus (outside of that segment). Once we have the potential distribution, we can find the fluid pressure at the centre of the disc from Eq.~\eqref{eq:enthalpy_eq},
\begin{equation}
p_{\mathrm{c}} = w_{\mathrm{c}}(W_{\mathrm{in}} - W_{\mathrm{c}})\left(\frac{\kappa}{\kappa - 1} + \frac{\lambda
}{\lambda
 - 1}\frac{1}{\beta_{\mathrm{m}_{\mathrm{c}}}}\right)^{-1},
\end{equation}
where $w_{\mathrm{c}}$ is the enthalpy density at the centre and
\begin{equation}
\label{eq:beta_eq}
\beta_{\mathrm{m}} = p/p_{\mathrm{m}},
\end{equation}
is the magnetisation parameter ($\beta_{\mathrm{m}_{\mathrm{c}}}$ being the magnetisation parameter at the centre of the disc). Using this definition, we can find the magnetic pressure at the centre,
\begin{equation}
p_{\mathrm{m_{\mathrm{c}}}} = p_{\mathrm{c}}/\beta_{\mathrm{m}_{\mathrm{c}}}\,.
\end{equation}
With both pressures known at the centre, fluid and magnetic, we can now find the constants $K$ and $K_{\mathrm{m}}$ using equations \eqref{eq:eos_fluid} and \eqref{eq:eos_mag}. Therefore, for a given inner radius of the disc $r_{\mathrm{in}}$ we can obtain the potential $W_{\mathrm{in}}$. We now finally have all the ingredients required to find the enthalpy density distribution (Eq.~\eqref{eq:enthalpy_eq}), the fluid pressure and magnetic pressure distributions (Eqs.~\eqref{eq:eos_fluid} and \eqref{eq:eos_mag}), and the rest-mass density distribution, which can be trivially obtained inverting the barotropic relation, $p = K \rho^{\kappa}$. We note that the fluid enthalpy density $w$ includes the rest-mass density $\rho$, and can thus be defined as $w=\rho h$ where $h$ is the specific enthalpy. The relativistic definition of this quantity is $h=1+\varepsilon+p/\rho$, where $\varepsilon$ is
the specific internal energy. From a thermodynamical point of view a non-relativistic fluid satisfies $h=1$. Therefore,
since we use poytropic equations of state (relating $p$ with either $w$ or $\rho$ in the same functional form) we are
implicitly assuming that $h=1$, i.e.~the discs we build in our procedure are non-relativistic from a thermodynamical point of view.

For the integration of Eq.~\eqref{eq:equatorial_pot} we use the composite Simpson's rule. It is important to use a very small integration step because the slope is very steep. In this work, we use a step $\Delta r = 10^{-6}$. Using the analytic, constant angular momentum case for comparison, we tested that 
a larger value of $\Delta r$ gives unacceptable accuracy losses. On the other hand, to integrate the ordinary differential equation \eqref{eq:F} we use a fourth-order Runge-Kutta method. Again, it is also important here to choose a suitable step of integration, especially at the outer end of the disc, because Eq.~\eqref{eq:F} diverges at the equatorial plane (the equipotential surfaces cross the equatorial plane perpendicularly). We show the proof of this statment in Appendix~\ref{div_partial_W}.

\begin{table}
\caption{List of models for $\beta_{\mathrm{m}_{\mathrm{c}}} = 10^{3}$. From left to right the columns report: the model name (with A, B and C standing for black hole spin $a = 0.5$, $0.9$ and $0.99$, respectively), the parameters of the angular momentum distribution $\beta$ and $\gamma$, the corresponding radii of the maximum fluid pressure and magnetic pressure, $r_{{\rm{max}}}$ and $r_{{\mathrm{m, max}}}$, the gravitational potential difference between the centre and the inner radius, $\Delta W\equiv W_{\rm c}-W_{\rm in}$, and the inner and outer radii of the discs, $r_{\mathrm{in}}$ and $r_{\mathrm{out}}$.}             
\label{table:1}      
\centering          
\begin{tabular}{c c c c  c c c c}
\hline\hline       
 & $\beta$ & $\gamma$ & $r_{\rm{max}}$ &  $r_{\mathrm{m, max}}$ & $\Delta W$               & $r_{\mathrm{in}}$ & $r_{\mathrm{out}}$ \\ 
 &              &                   &                          &                                        & $(\times 10^{-2})$     &                              &  \\
\hline           
A1 & $0.0$ & $0.0$ & $7.15$ &  $8.14$  & $-6.35$ & $2.91$ & -- \\ 
A2 & $0.5$ & $0.5$ & $7.15$ &  $7.65$  & $-2.27$ & $3.20$ & $11.8$\\ 
A3 & $0.9$ & $0.9$ & $7.15$ &  $7.45$  & $-0.30$ & $3.70$ &  $9.58$\\ 
A4 & $0.0$ & $0.9$ & $7.15$ &  $8.32$  & $-6.35$ & $2.91$ & $9.68$\\ 
A5 & $0.9$ & $0.0$ & $7.15$ &  $7.28$  & $-0.30$ & $3.70$ & --\\ 
 \hline 
B1 & $0.0$ & $0.0$ & $3.59$ &  $3.98$  & $-12.9$ & $1.73$ & -- \\ 
B2 & $0.5$ & $0.5$ & $3.59$ &  $3.83$  & $-4.32$ & $1.86$ & $5.35$\\ 
B3 & $0.9$ & $0.9$ & $3.59$ &  $3.73$  & $-0.54$ & $2.08$ & $4.52$\\ 
B4 & $0.0$ & $0.9$ & $3.59$ &  $3.85$  & $-12.9$ & $1.73$ & $4.54$\\ 
B5 & $0.9$ & $0.0$ & $3.59$ &  $3.81$  & $-0.54$ & $2.08$ & -- \\  
 \hline 
C1 & $0.0$ & $0.0$ & $1.98$ &  $2.28$  & $-24.6$ & $1.21$ & -- \\ 
C2 & $0.5$ & $0.5$ & $1.98$ &  $2.09$  & $-7.34$ & $1.26$ & $2.50$\\ 
C3 & $0.9$ & $0.9$ & $1.98$ &  $2.04$  & $-0.85$ & $1.35$ & $2.25$\\ 
C4 & $0.0$ & $0.9$ & $1.98$ &  $2.10$  & $-24.6$ & $1.21$ & $2.26$\\ 
C5 & $0.9$ & $0.0$ & $1.98$ &  $2.08$  & $-0.85$ & $1.35$ & --\\ 
\hline      
\end{tabular}
\end{table}

\begin{table}
\caption{List of models for $\beta_{\mathrm{m}_{\mathrm{c}}} = 1$. The naming of the models and the parameters reported in the columns are as in Table~\ref{table:1}.}             
\label{table:2}      
\centering          
\begin{tabular}{c c c c  c c c c}
\hline\hline       
 & $\beta$ & $\gamma$ & $r_{\rm{max}}$ &  $r_{\mathrm{m, max}}$ & $\Delta W$               & $r_{\mathrm{in}}$ & $r_{\mathrm{out}}$ \\ 
 &              &                   &                          &                                        & $(\times 10^{-2})$     &                              &  \\
\hline           
A1 & $0.0$ & $0.0$ & $6.00$ &  $6.49$  & $-6.35$ & $2.91$ & -- \\ 
A2 & $0.5$ & $0.5$ & $6.25$ &  $6.66$  & $-2.27$ & $3.20$ & $11.8$\\ 
A3 & $0.9$ & $0.9$ & $6.50$ &  $6.80$  & $-0.30$ & $3.70$ &  $9.58$\\ 
A4 & $0.0$ & $0.9$ & $6.00$ &  $6.49$  & $-6.35$ & $2.91$ & $9.68$\\ 
A5 & $0.9$ & $0.0$ & $6.50$ &  $6.80$  & $-0.30$ & $3.70$ & --\\ 
 \hline 
B1 & $0.0$ & $0.0$ & $3.02$ &  $3.26$  & $-12.9$ & $1.73$ & -- \\ 
B2 & $0.5$ & $0.5$ & $3.16$ &  $3.35$  & $-4.32$ & $1.86$ & $5.35$\\
B3 & $0.9$ & $0.9$ & $3.29$ &  $3.43$  & $-0.54$ & $2.08$ & $4.52$\\ 
B4 & $0.0$ & $0.9$ & $3.02$ &  $3.26$  & $-12.9$ & $1.73$ & $4.54$\\  
B5 & $0.9$ & $0.0$ & $3.29$ &  $3.43$  & $-0.54$ & $2.08$ & -- \\ 
 \hline 
C1 & $0.0$ & $0.0$ & $1.68$ &  $1.80$  & $-0.246$ & $1.21$ & -- \\ 
C2 & $0.5$ & $0.5$ & $1.77$ &  $1.86$  & $-7.34$ & $1.26$ & $2.50$\\  
C3 & $0.9$ & $0.9$ & $1.84$ &  $1.91$  & $-0.85$ & $1.35$ & $2.25$\\ 
C4 & $0.0$ & $0.9$ & $1.68$ &  $1.80$  & $-24.6$ & $1.21$ & $2.26$\\ 
C5 & $0.9$ & $0.0$ & $1.84$ &  $1.91$  & $-0.85$ & $1.35$ & --\\ 
\hline      
\end{tabular}
\end{table}

\section{Results}
\label{results}

To reduce the number of free parameters to build the initial models we do as in~\citet{Komissarov:2006} and fix the values of the equation of state exponents, $\kappa$ and $\lambda$, and of the enthalpy density at the disc centre, $w_{\mathrm{c}}$. More precisely, we choose $\kappa = \lambda = 4/3$ and $w_{\mathrm{c}} = 1$. This still leaves us with five parameters to control the size, shape, thickness, and magnetisation of the disc, namely the magnetisation parameter $\beta_{\mathrm{m}_{\mathrm{c}}}$, the parameters of the angular momentum distribution $\beta$ and $\gamma$, the black hole spin parameter $a$, and the inner radius of the disc $r_{\mathrm{in}}$. We build a series of 45 models, whose main features are summarised in Table~\ref{table:1} (for $\beta_{\mathrm{m}_{\mathrm{c}}} = 10^{3}$, i.e.~models where the effects of the magnetisation are unimportant), Table~\ref{table:2} ($\beta_{\mathrm{m}_{\mathrm{c}}} = 1$, i.e.~equipartition models), and Table~\ref{table:3} ($\beta_{\mathrm{m}_{\mathrm{c}}} = 10^{-3}$, i.e.~highly-magnetised models). We note that we can build models with very small values of $\beta_{\mathrm{m}_{\mathrm{c}}}$, such as  e.g.~$\beta_{\mathrm{m}_{\mathrm{c}}} = 10^{-20}$ (extremely high magnetisation) without encountering numerical difficulties. No qualitative differences in the structure of the discs are found once $\beta_{\mathrm{m}_{\mathrm{c}}}$ becomes smaller than a sufficiently small value, about $\beta_{\mathrm{m}_{\mathrm{c}}}=10^{-3}$. This is the reason why we choose that particular value as our lower limit in the figures and tables of the manuscript.

We note in particular that the radii of the maximum fluid pressure and magnetic pressure, $r_{{\rm{max}}}$ and $r_{{\mathrm{m, max}}}$, respectively, for the same value of $\beta_{\mathrm{m}_{\mathrm{c}}}$, never coincide. This also reflects the fact that constant fluid pressure surfaces do not coincide with constant magnetic pressure surfaces.

We start by assessing our procedure by first building an extra disc model that can be directly compared with one of the two models in~\citet{Komissarov:2006}. This is shown in Figure~\ref{komissarov}, which corresponds to the same constant specific angular momentum model A presented by~\citet{Komissarov:2006}. The parameters of this model are $a=0.9$, $\beta_{\mathrm{m}_{\mathrm{c}}}=0.1$, and $l=2.8$.
The visual comparison shows that our approach can reproduce those previous results with good agreement. The rest-mass density distribution in the $(r\sin\theta,r\cos\theta)$ plane shown in the left panel of Fig.~\ref{komissarov} is remarkably similar to that shown in the left panel of Fig.~2 in~\citet{Komissarov:2006}. Not only the morphology of both models is nearly identical but also the range of variation of the rest-mass density and the location of the disc centre agree well in both cases. This can be most clearly seen in the middle and right panels of Fig.~\ref{komissarov} which show, repectively, the angular profile at $r_{\rm c}$ and the radial profile at $\theta=\pi/2$. The middle figure can be directly compared with Fig.~3 of~\citet{Komissarov:2006}. It is relevant to mention that very small changes in the location of the inner radius $r_{\mathrm{in}}$ have a significant effect on the maximum value of the rest-mass density, which explains the small differences between our figures and the ones presented in~\citet{Komissarov:2006}.

\begin{table}
\caption{List of models for $\beta_{\mathrm{m}_{\mathrm{c}}} = 10^{-3}$. The naming of the models and the parameters reported in the columns are as in Table~\ref{table:1}.}             
\label{table:3}      
\centering          
\begin{tabular}{c c c c  c c c c}
\hline\hline       
 & $\beta$ & $\gamma$ & $r_{\rm{max}}$ &  $r_{\mathrm{m, max}}$ & $\Delta W$               & $r_{\mathrm{in}}$ & $r_{\mathrm{out}}$ \\ 
 &              &                   &                          &                                        & $(\times 10^{-2})$     &                              &  \\
\hline           
A1 & $0.0$ & $0.0$ & $5.11$ &  $5.51$  & $-6.35$ & $2.91$ & -- \\ 
A2 & $0.5$ & $0.5$ & $5.55$ &  $5.80$  & $-2.27$ & $3.20$ & $11.8$\\ 
A3 & $0.9$ & $0.9$ & $5.95$ &  $6.15$  & $-0.30$ & $3.70$ &  $9.58$\\ 
A4 & $0.0$ & $0.9$ & $5.21$ &  $5.51$  & $-6.35$ & $2.91$ & $9.68$\\  
A5 & $0.9$ & $0.0$ & $5.90$ &  $6.10$  & $-0.30$ & $3.70$ & --\\
 \hline 
B1 & $0.0$ & $0.0$ & $2.53$ &  $2.73$  & $-12.9$ & $1.73$ & -- \\ 
B2 & $0.5$ & $0.5$ & $2.81$ &  $2.91$  & $-4.32$ & $1.86$ & $5.35$\\
B3 & $0.9$ & $0.9$ & $3.03$ &  $3.13$  & $-0.54$ & $2.08$ & $4.52$\\  
B4 & $0.0$ & $0.9$ & $2.65$ &  $2.80$  & $-12.9$ & $1.73$ & $4.54$\\ 
B5 & $0.9$ & $0.0$ & $3.03$ &  $3.13$  & $-0.54$ & $2.08$ & -- \\  
 \hline 
C1 & $0.0$ & $0.0$ & $1.46$ &  $1.51$  & $-24.6$ & $1.21$ & -- \\ 
C2 & $0.5$ & $0.5$ & $1.56$ &  $1.61$  & $-7.34$ & $1.26$ & $2.50$\\ 
C3 & $0.9$ & $0.9$ & $1.70$ &  $1.75$  & $-0.85$ & $1.35$ & $2.25$\\ 
C4 & $0.0$ & $0.9$ & $1.50$ &  $1.57$  & $-24.6$ & $1.21$ & $2.26$\\ 
C5 & $0.9$ & $0.0$ & $1.72$ &  $1.77$  & $-0.85$ & $1.35$ & --\\ 
\hline      
\end{tabular}
\end{table}

\begin{figure*}
\centering
\includegraphics[scale=0.14]{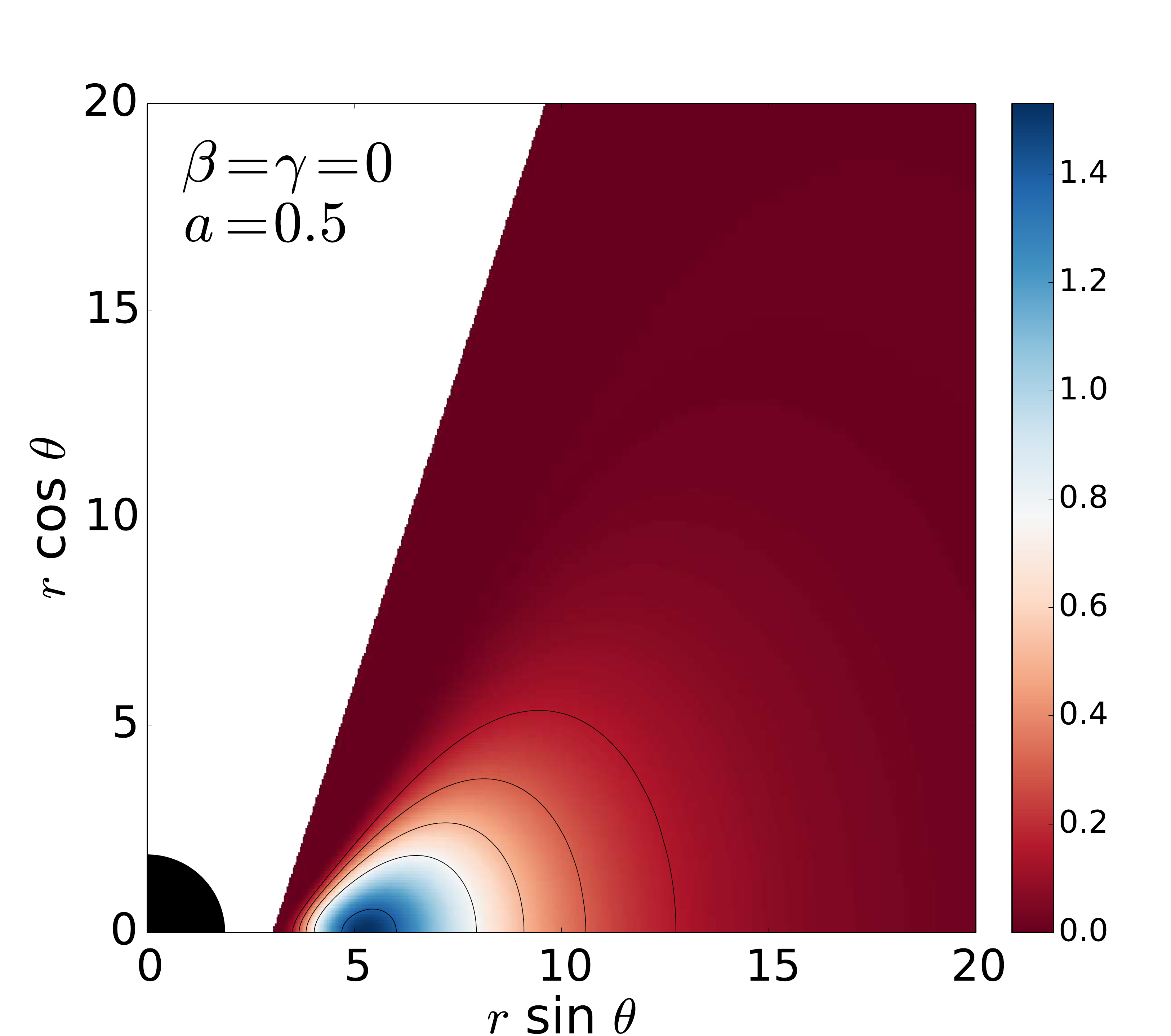}
\hspace{-0.3cm}
\includegraphics[scale=0.14]{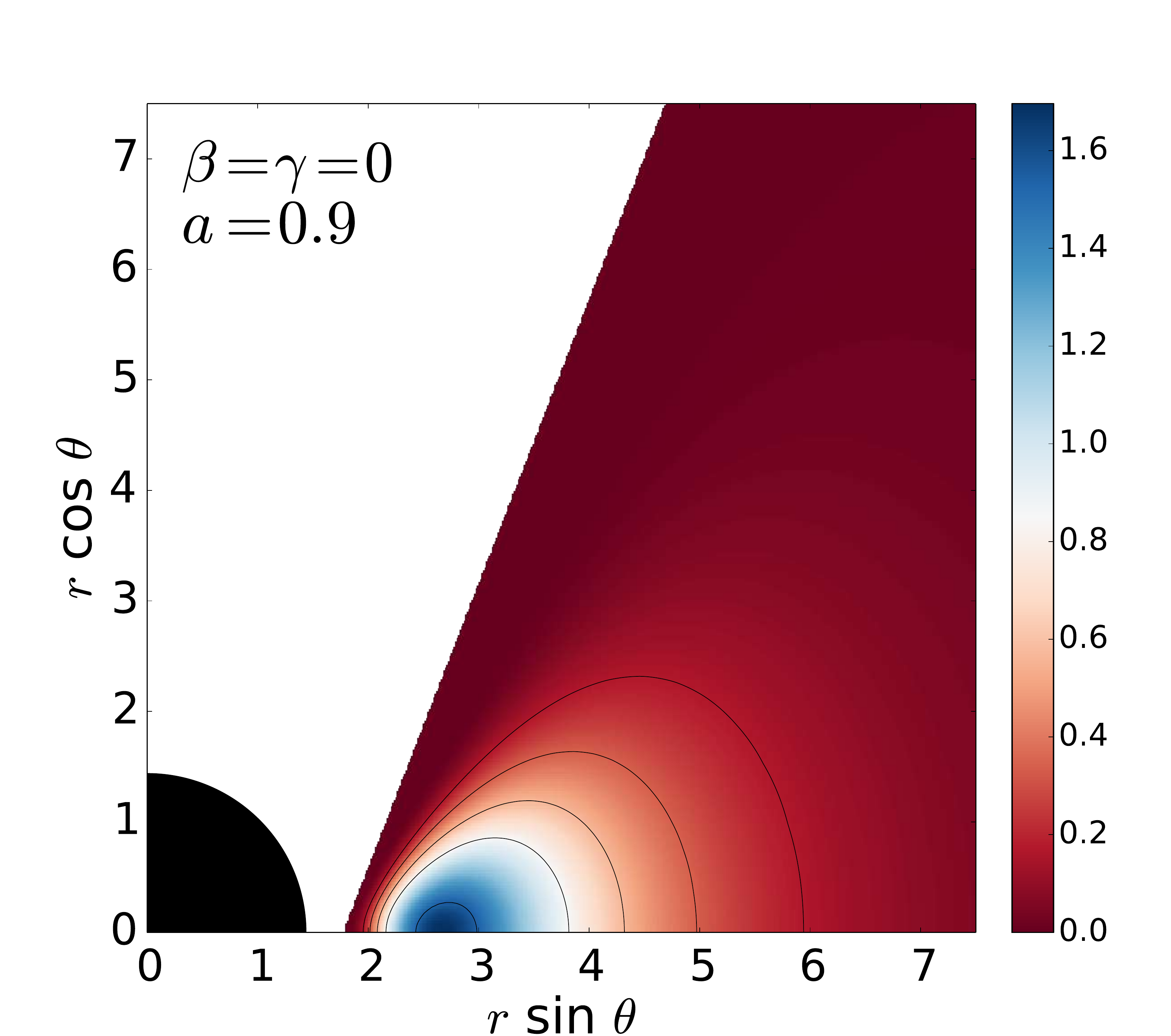}
\hspace{-0.2cm}
\includegraphics[scale=0.14]{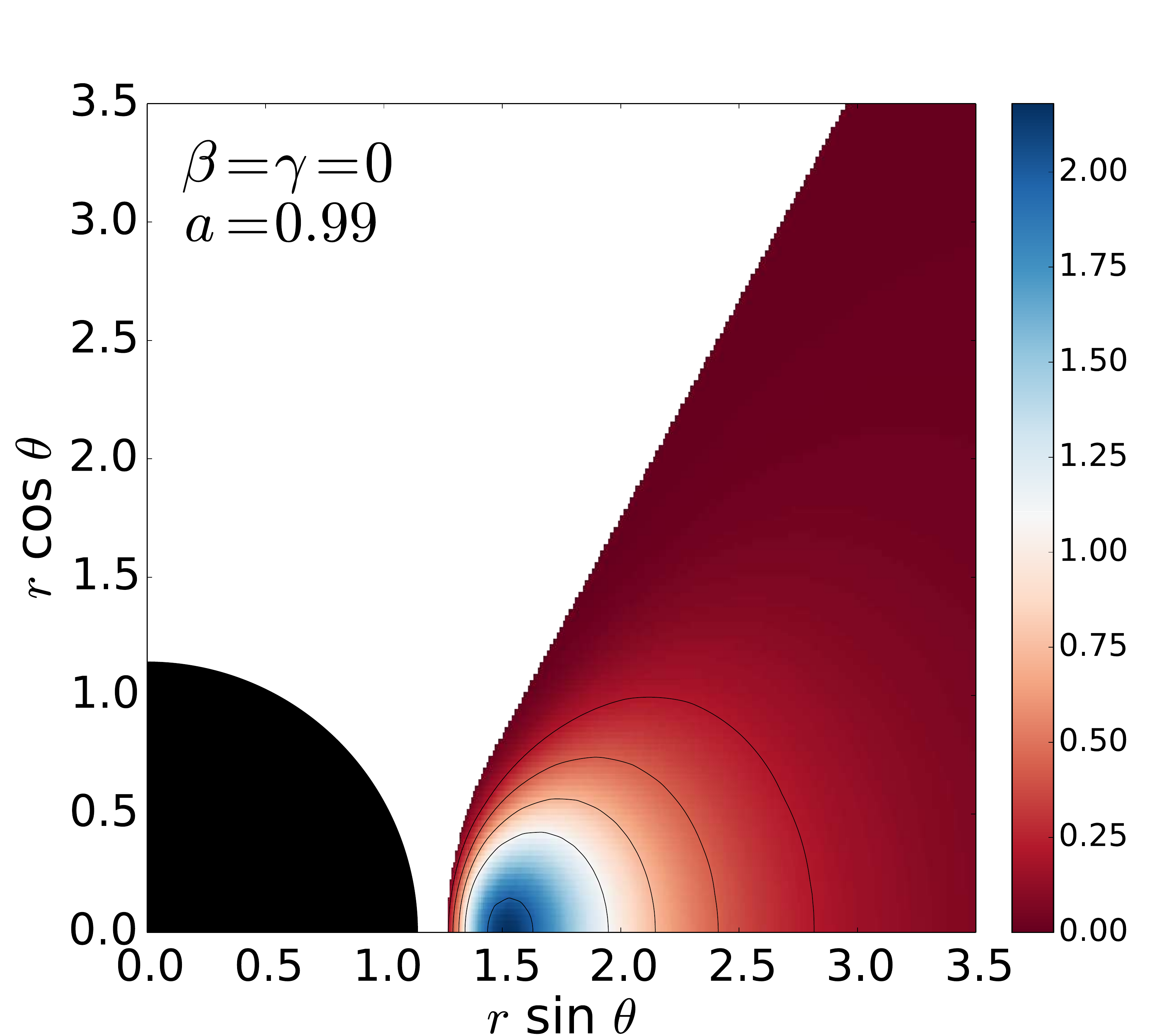}
\\
\includegraphics[scale=0.14]{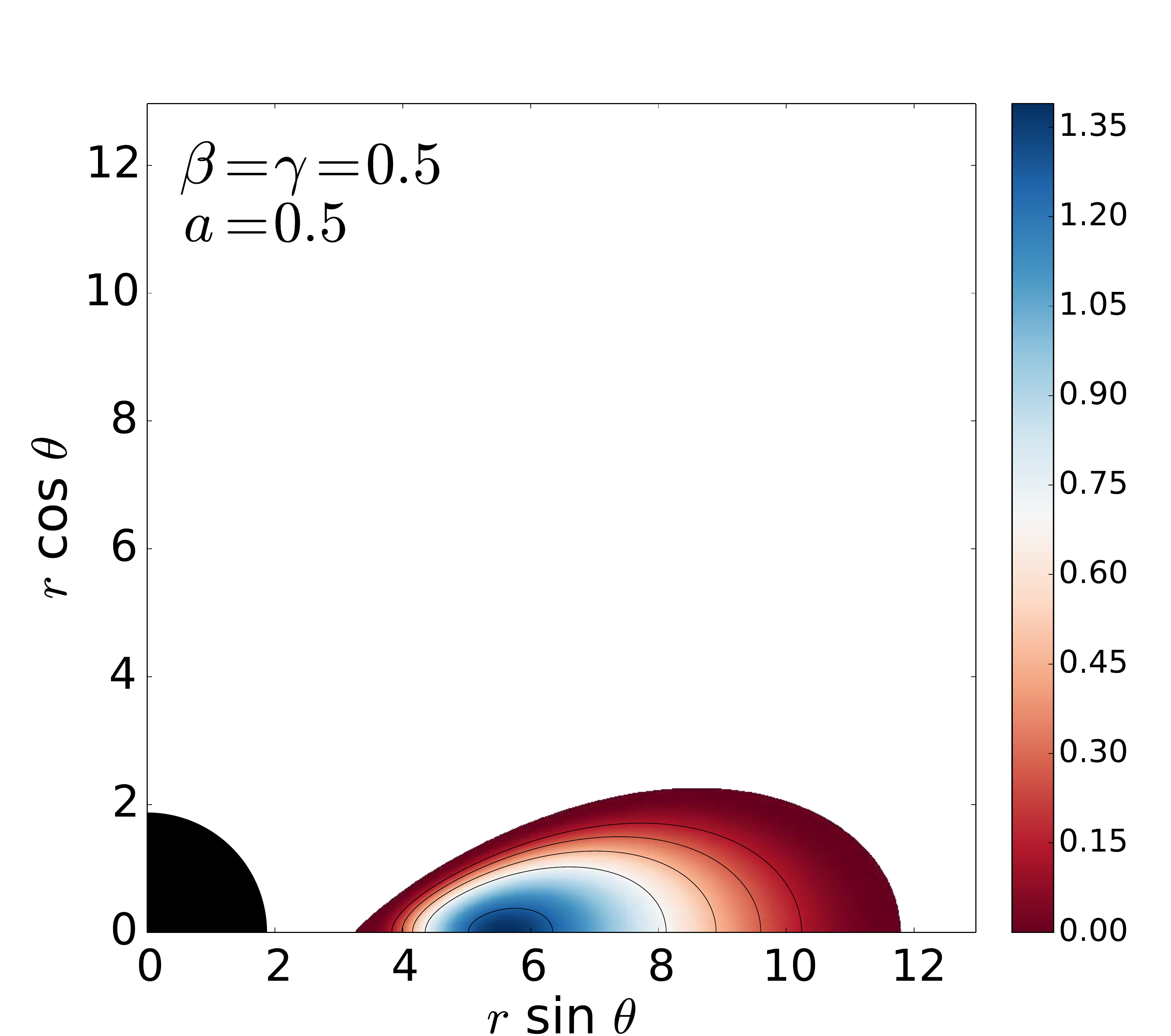}
\hspace{-0.3cm}
\includegraphics[scale=0.14]{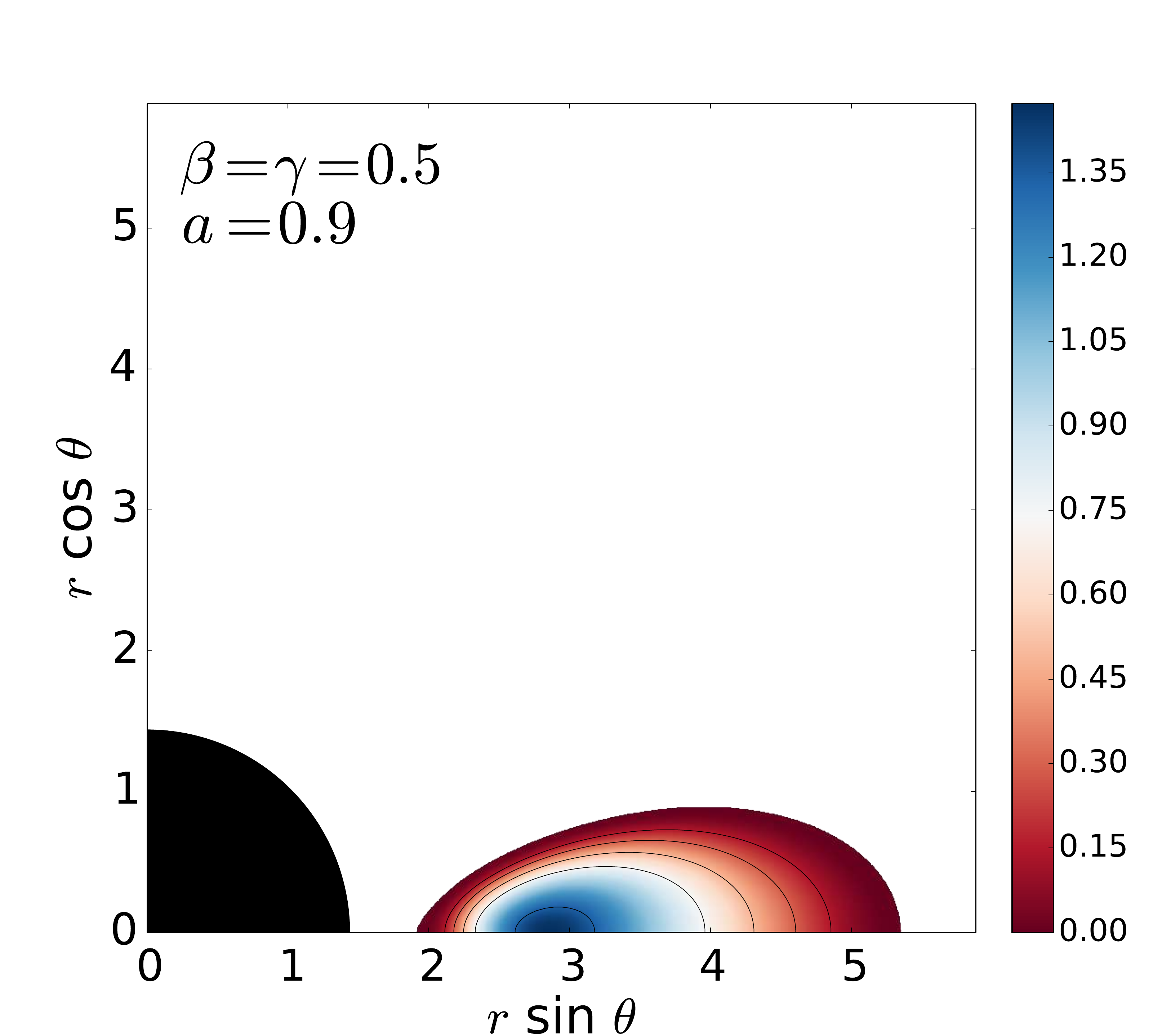}
\hspace{-0.2cm}
\includegraphics[scale=0.14]{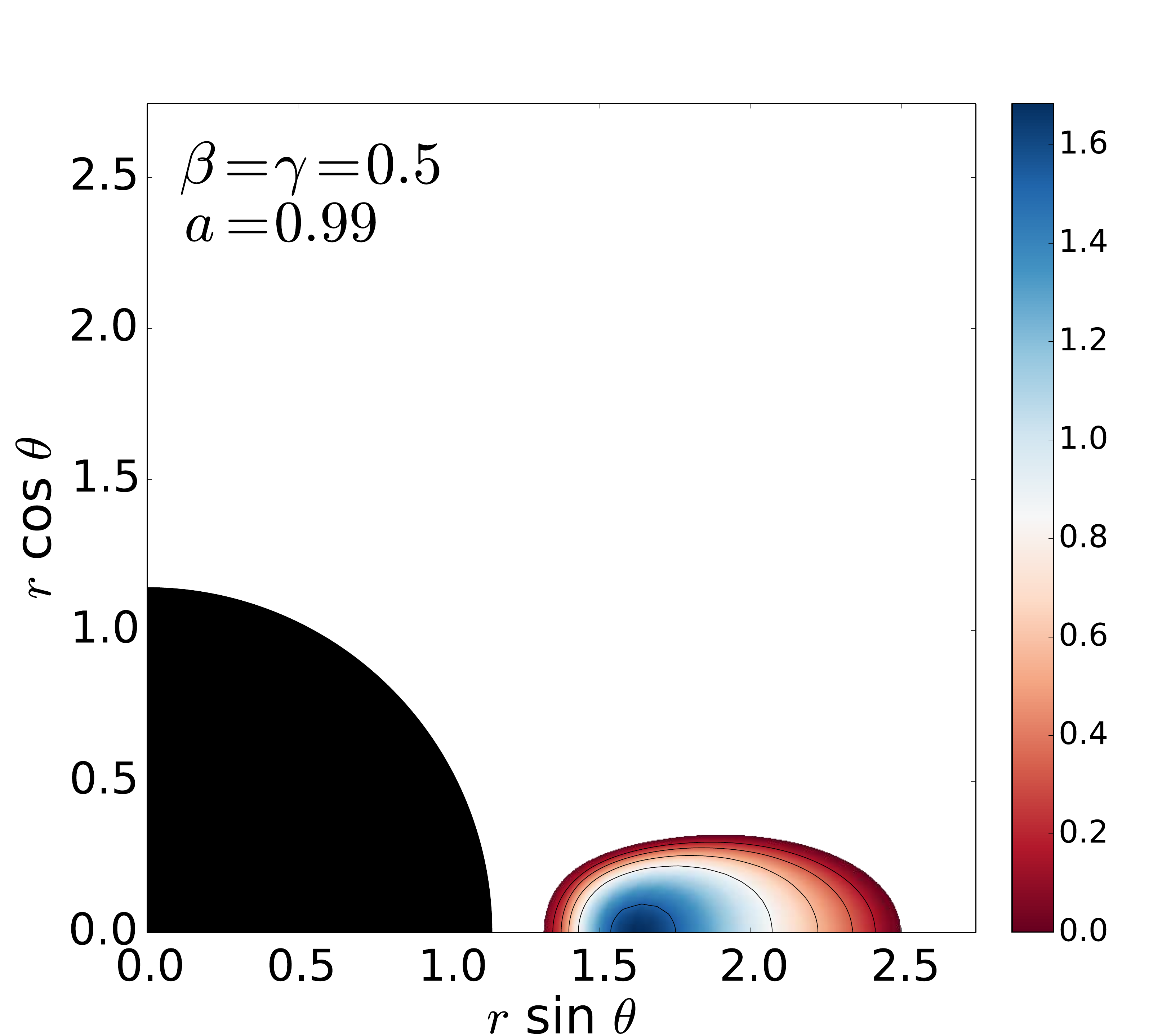}
\\
\includegraphics[scale=0.14]{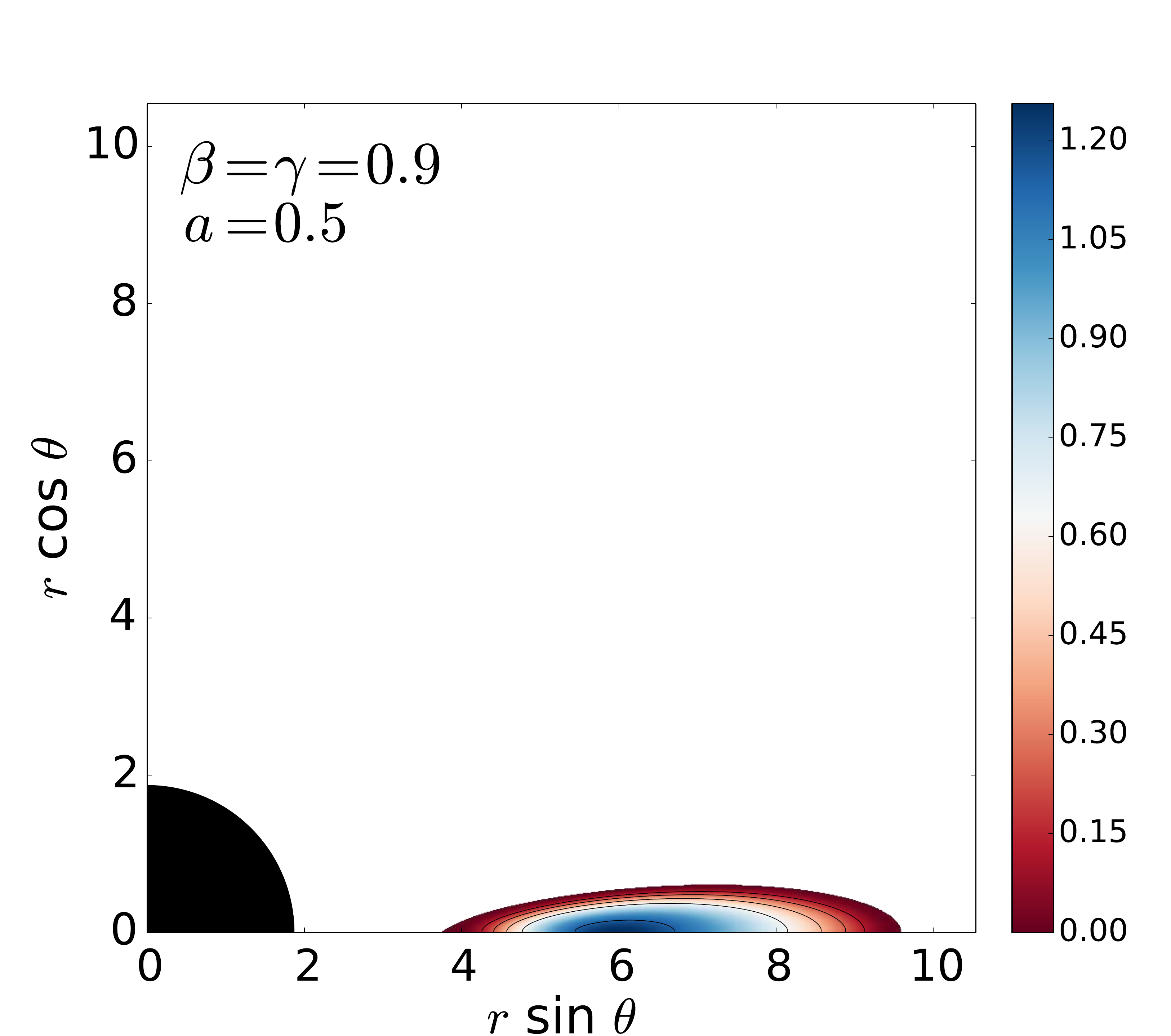}
\hspace{-0.3cm}
\includegraphics[scale=0.14]{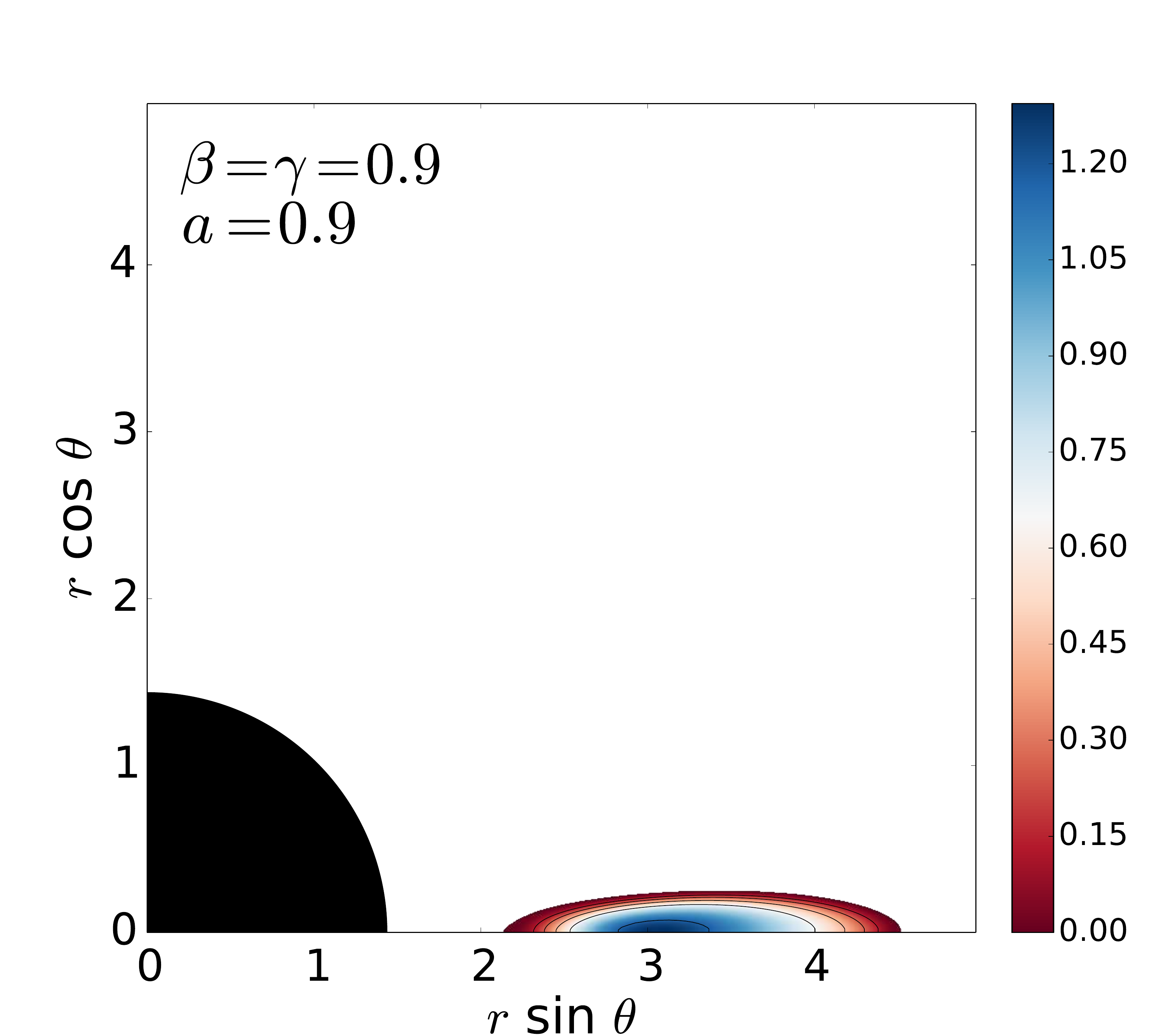}
\hspace{-0.2cm}
\includegraphics[scale=0.14]{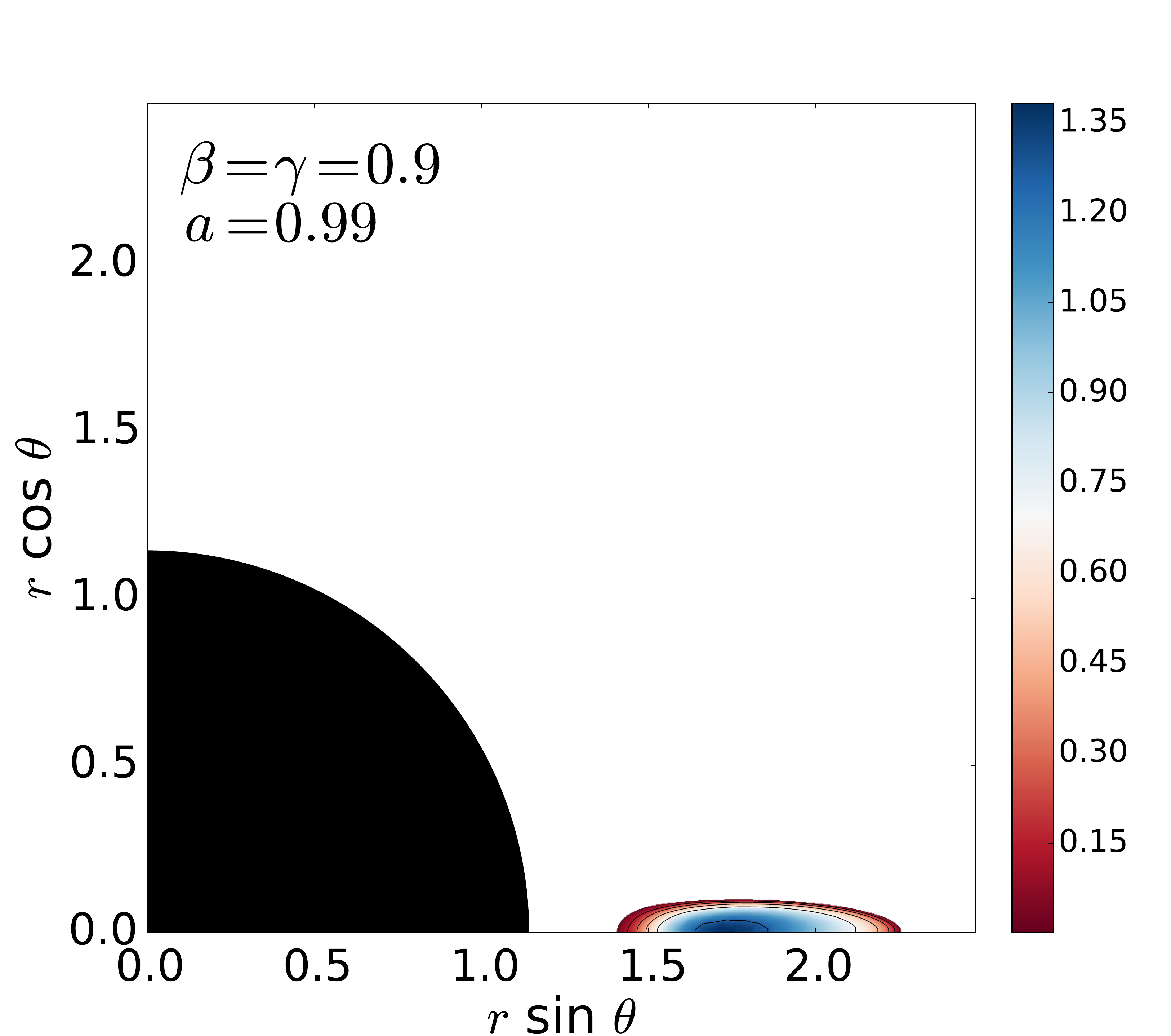}
\\
\includegraphics[scale=0.14]{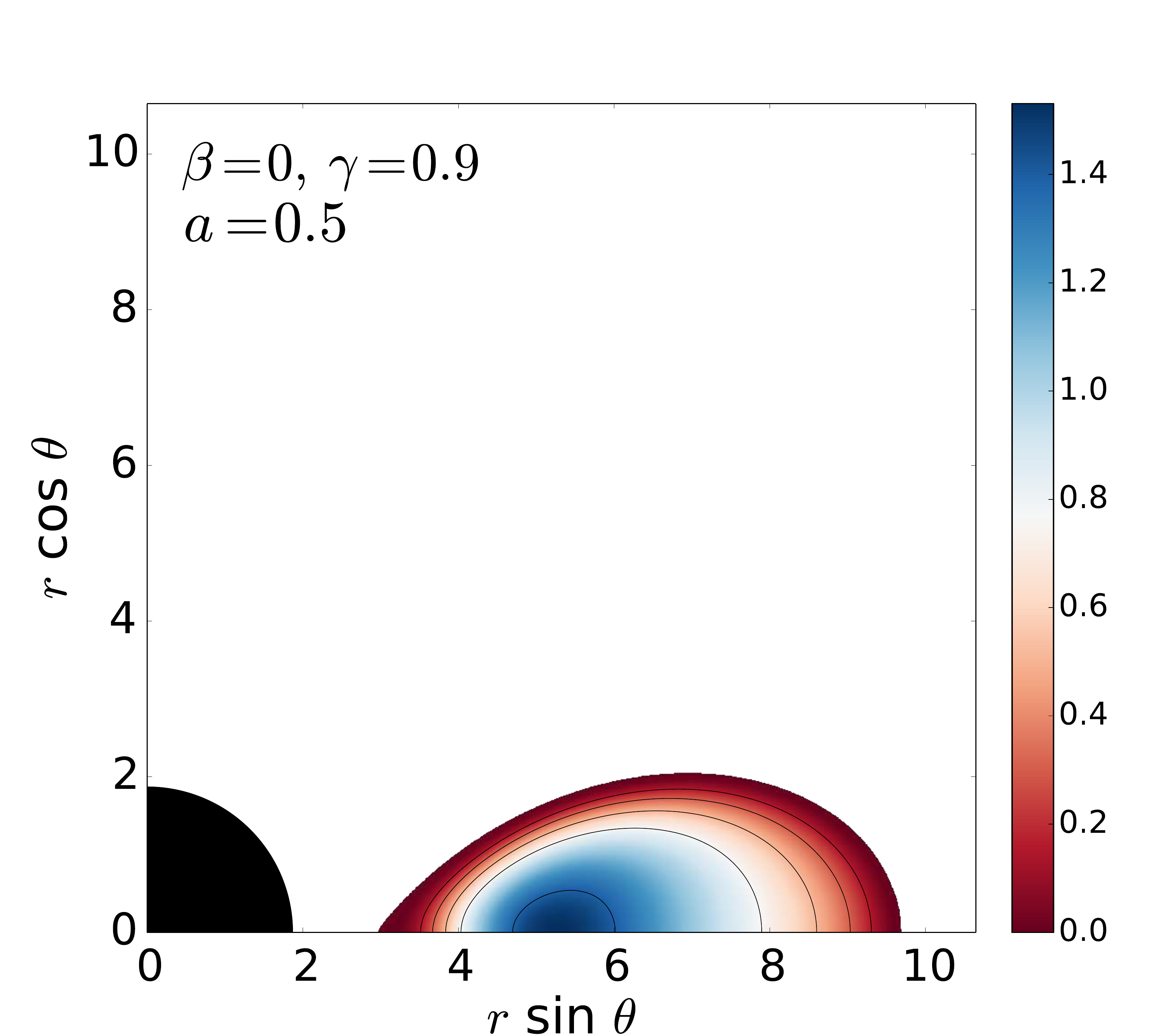}
\hspace{-0.3cm}
\includegraphics[scale=0.14]{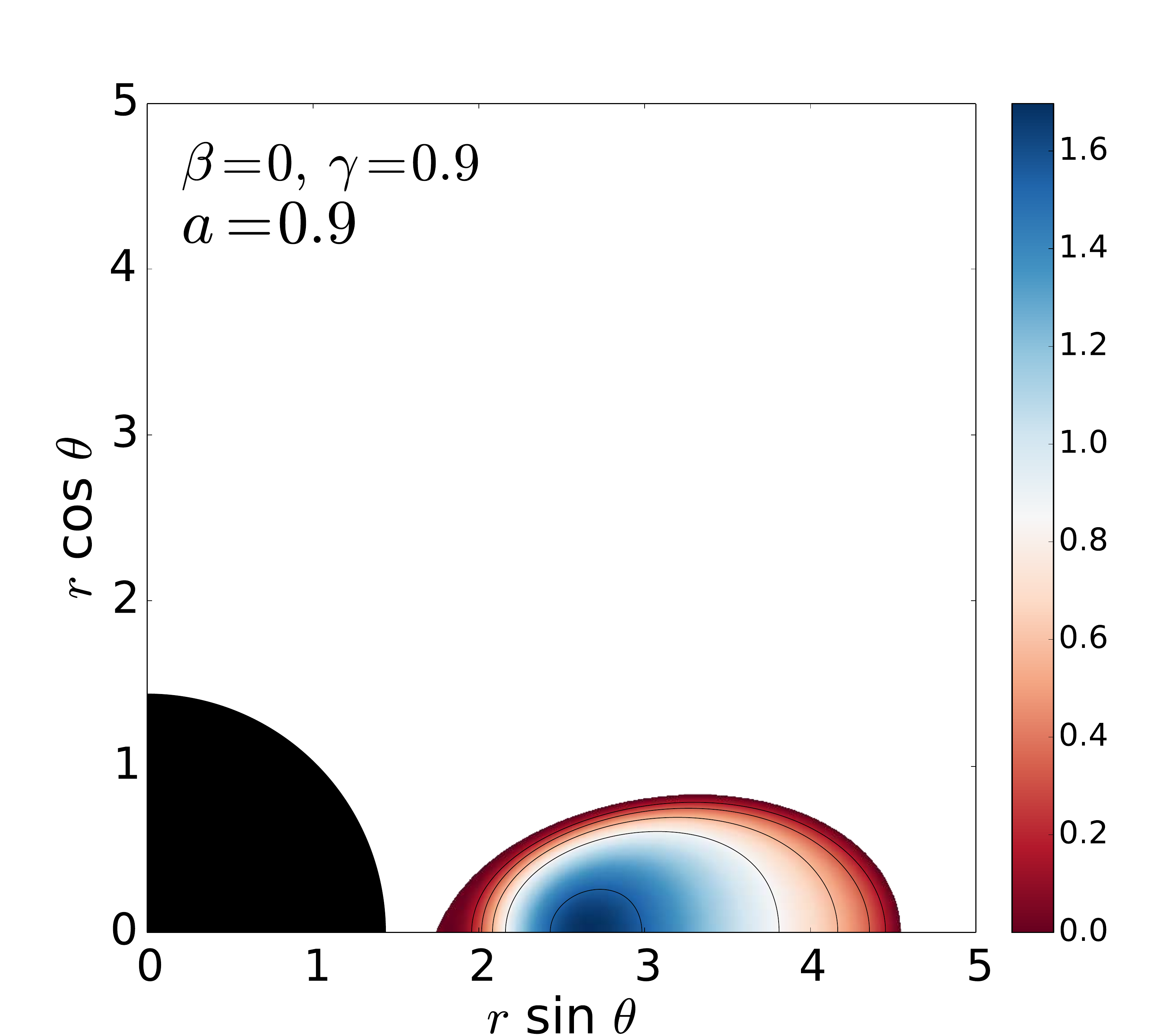}
\hspace{-0.2cm}
\includegraphics[scale=0.14]{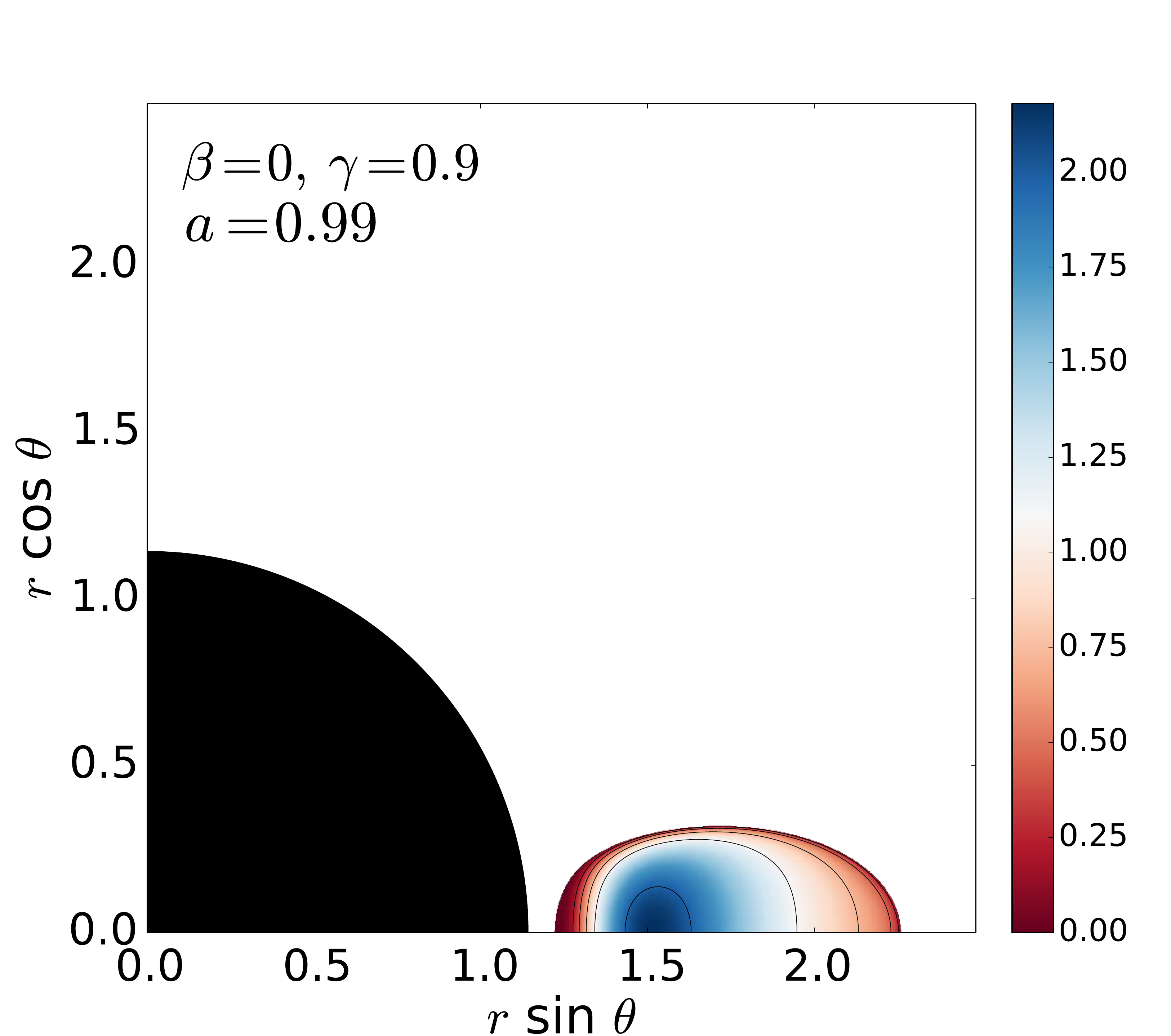}
\\
\includegraphics[scale=0.14]{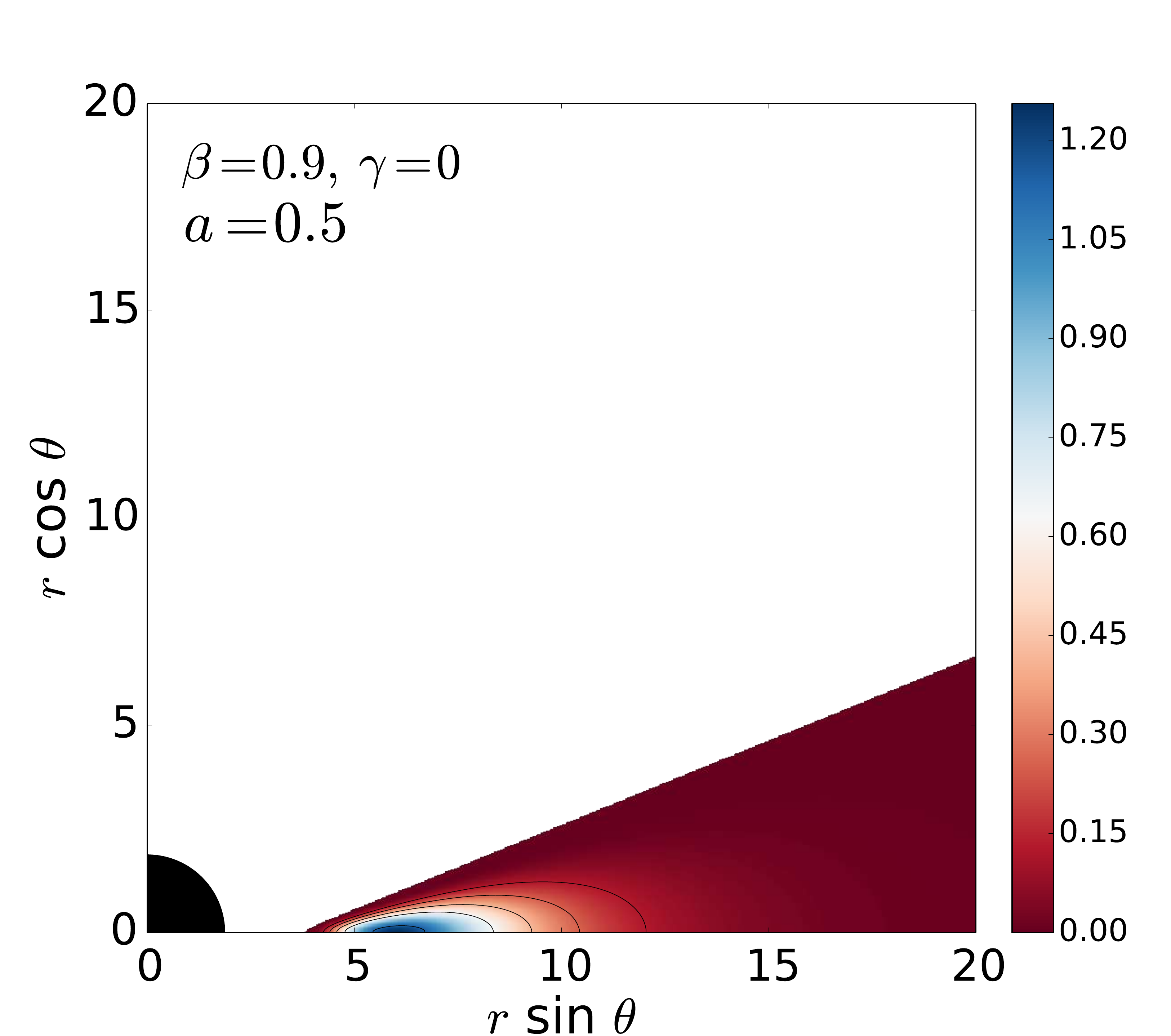}
\hspace{-0.3cm}
\includegraphics[scale=0.14]{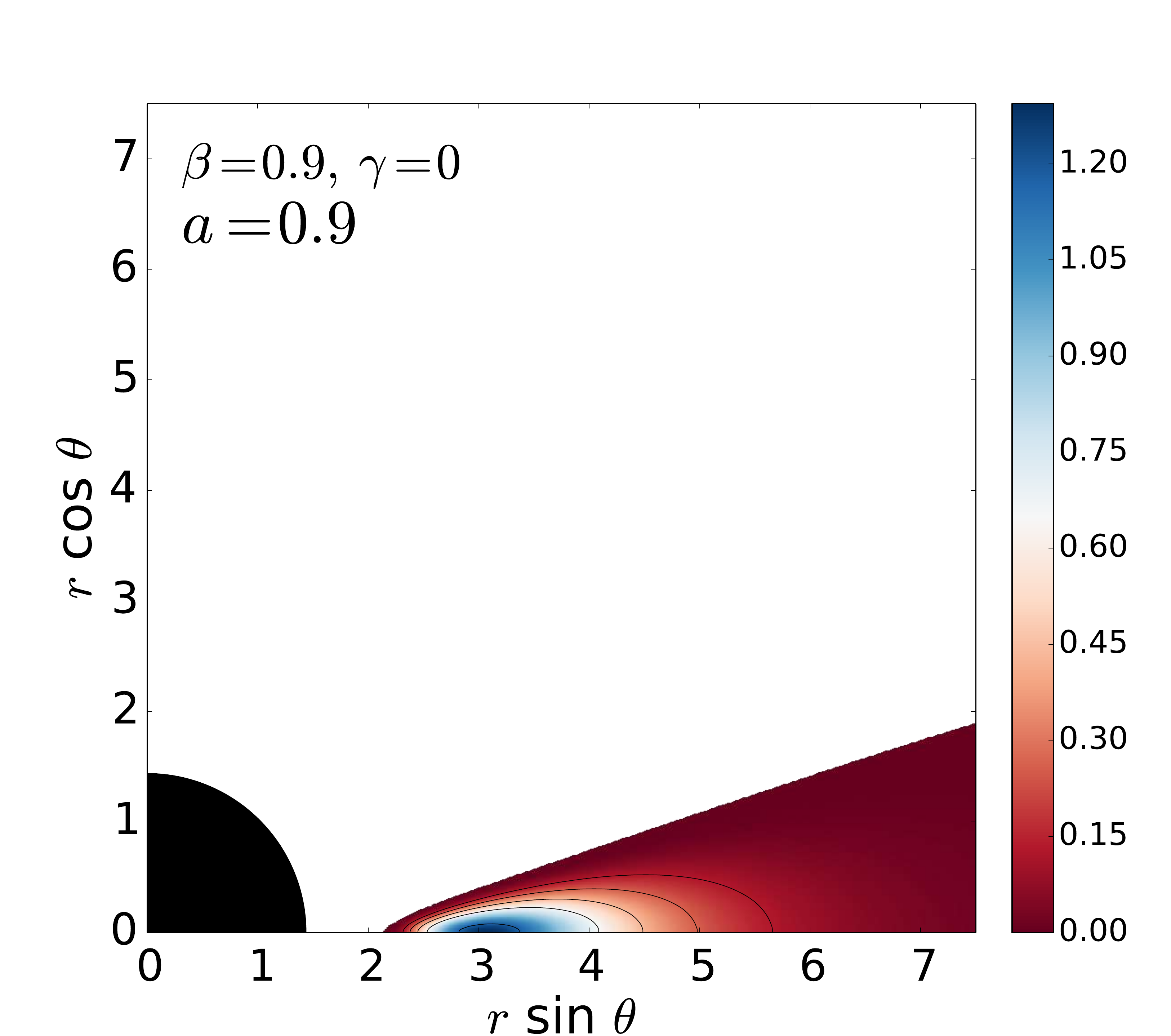}
\hspace{-0.2cm}
\includegraphics[scale=0.14]{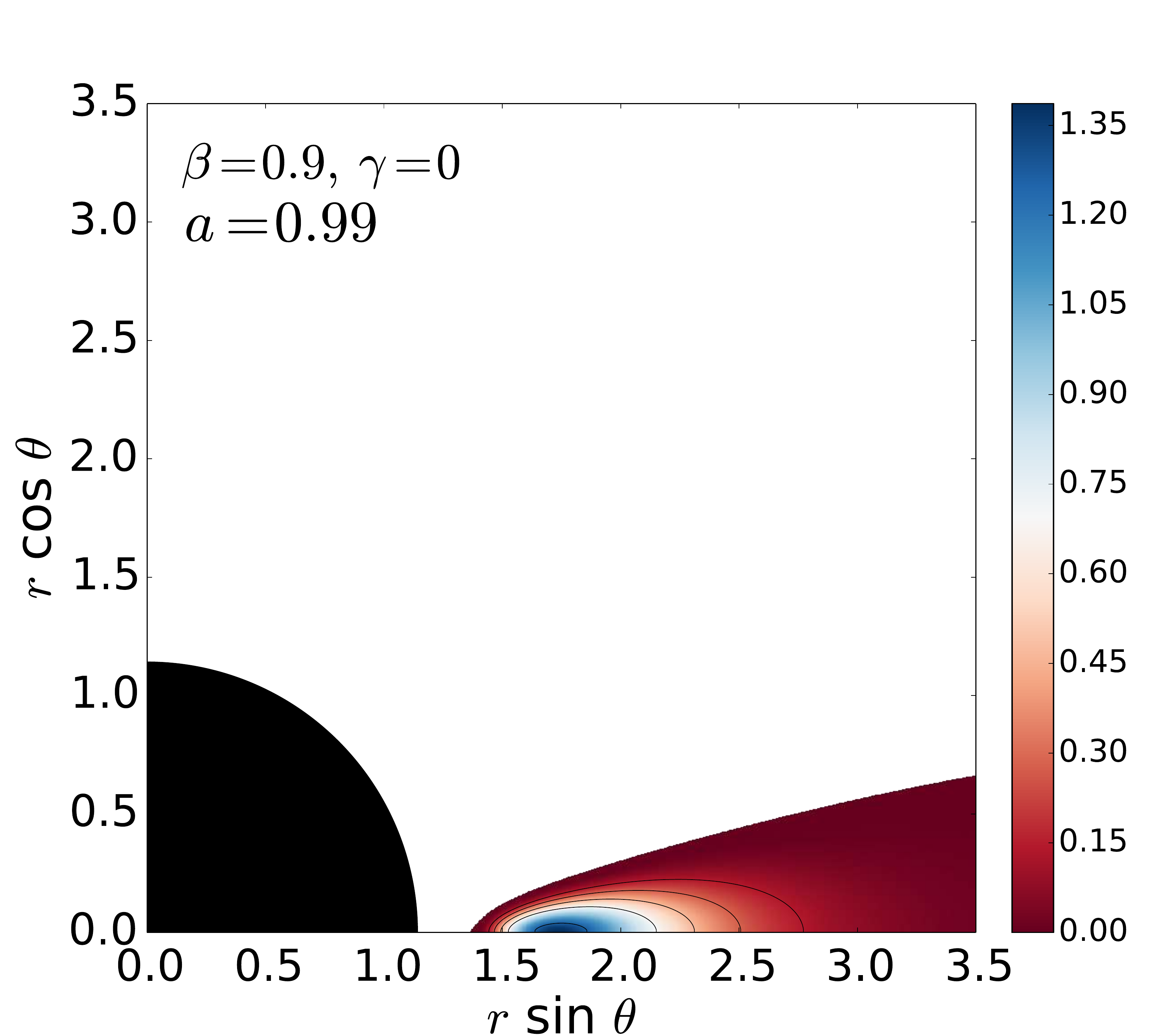}
\caption{Isodensity distributions for all models of Table~\ref{table:3}. From left to right the columns correspond to increasing values of the black hole spin: $a=0.5$, 0.9 and 0.99. From top to bottom the rows correspond to the following parameter combinations: a) $\gamma=\beta=0$; b) $\gamma=\beta=0.5$; c) $\gamma=\beta=0.9$; d) $\gamma=0.9$, $\beta=0$; e) $\gamma=0$, $\beta=0.9$. Note that the spatial scale of the plots differs as it has been chosen to facilitate the visualization of the discs.}
\label{models}
\end{figure*}

\begin{figure*}[t]
\centering
\includegraphics[scale=0.14]{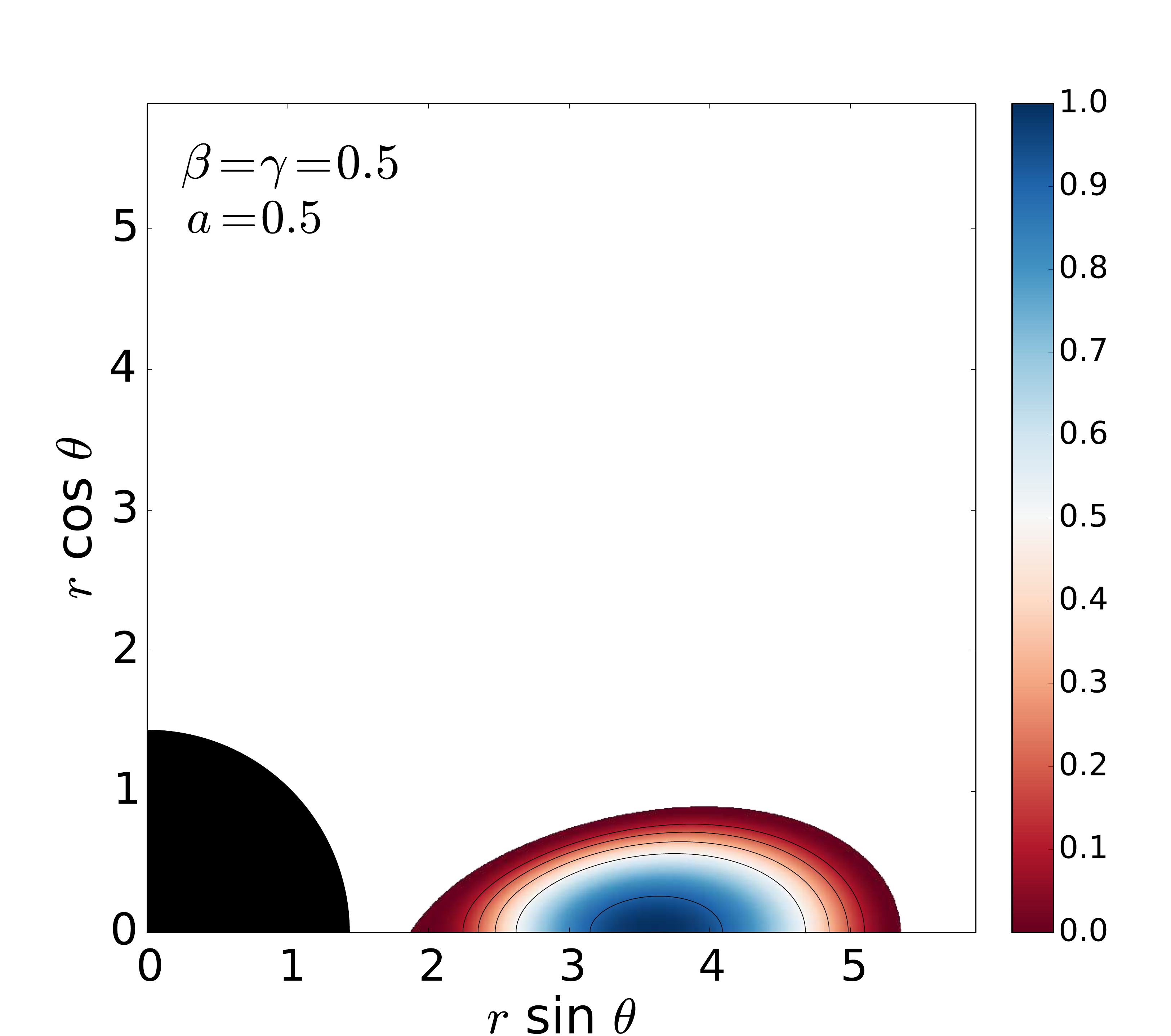}
\hspace{-0.3cm}
\includegraphics[scale=0.14]{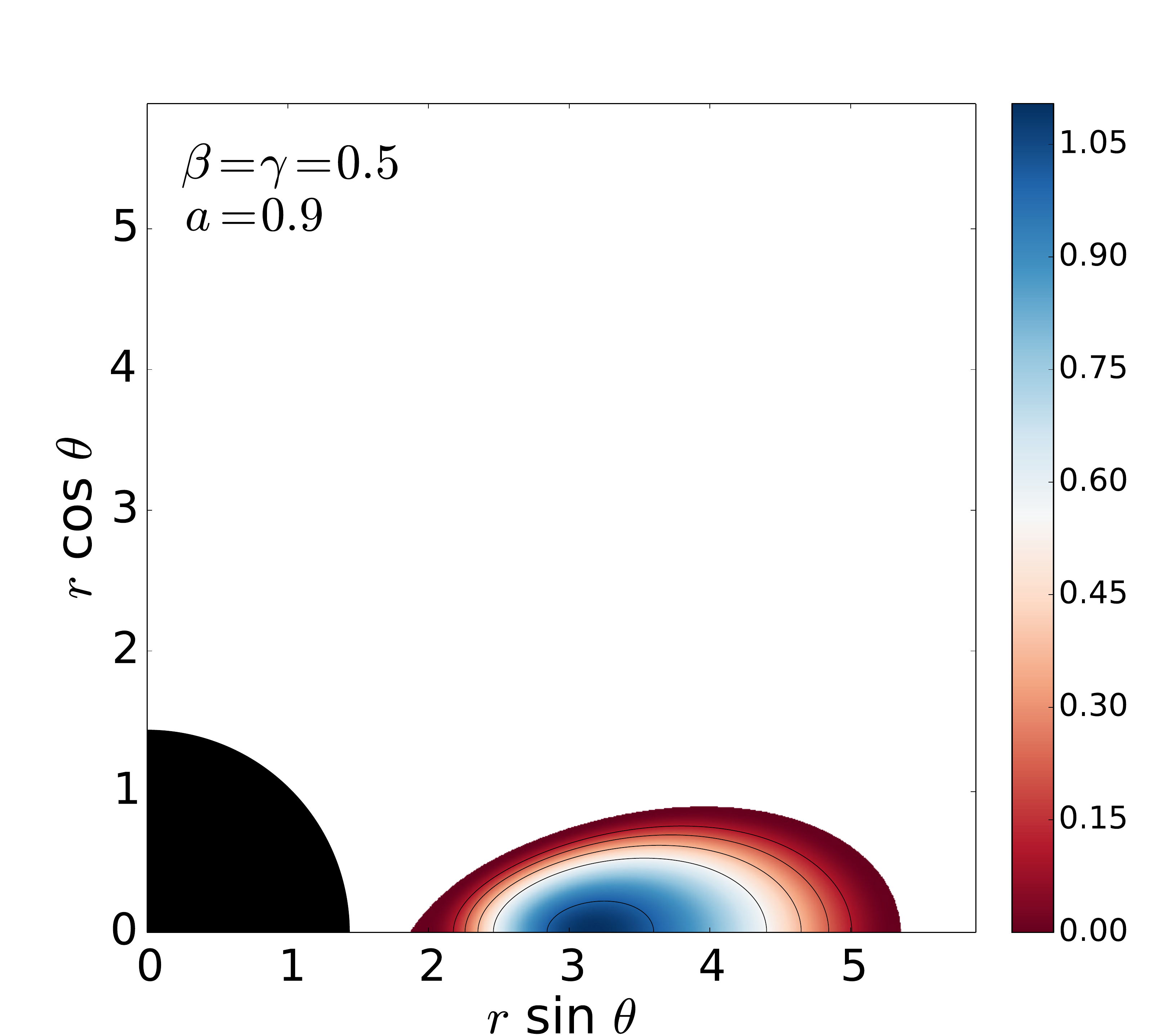}
\hspace{-0.2cm}
\includegraphics[scale=0.14]{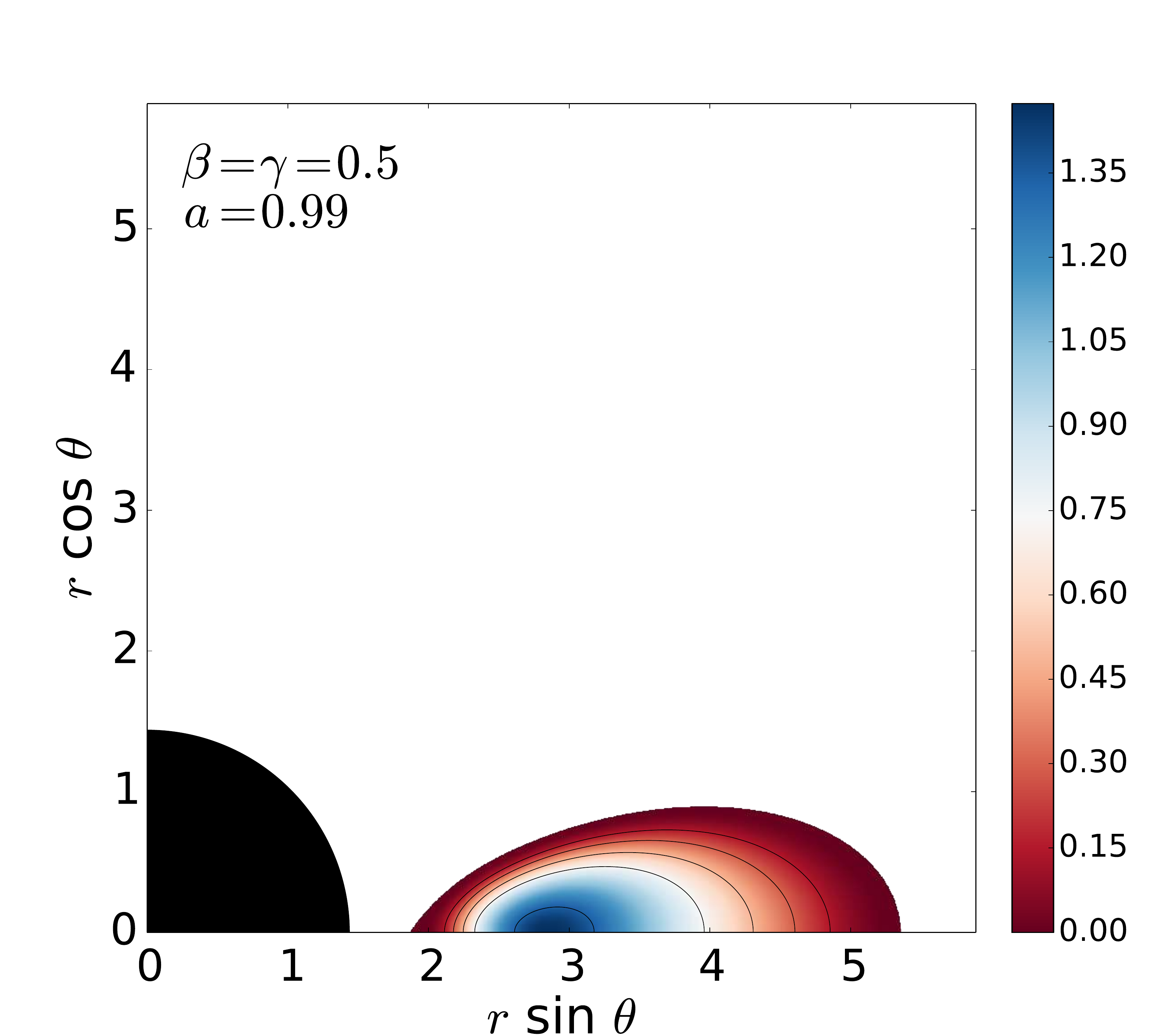}
\caption{Effects of the magnetisation in the structure of the disc. From left to right the values are $\beta_{\mathrm{m}_{\mathrm{c}}}
=10^3$, 1, and $10^{-3}$.}
\label{magnetisation}%
\end{figure*}

A representative sample of the isodensity surfaces for some of our models appears in Figure~\ref{models}. We plot, in particular, the models of Table~\ref{table:3}, for which the magnetisation is highest ($\beta_{\mathrm{m}_{\mathrm{c}}}=10^{-3}$). From left to right the columns of this figure correspond to increasing values of the black hole spin, namely, $a=0.5$, 0.9 and 0.99, while from top to bottom the rows correspond to different combinations of the $\gamma$ and $\beta$ parameters that characterize the ansatz for the angular momentum distribution of~\citet{Qian:2009} (the particular values are indicated in the caption of the figure).  A rapid inspection shows that the structure of the discs noticeable changes when the parameters change. Notice that the spatial scale in all of the plots of this figure has been chosen so as to facilitate the visualization of the discs (which can be fairly small in some cases) and, as such, is different in each plot. A typical Polish doughnut is represented by the model in the top-left panel. As the black hole spin increases, the discs are closer to the black hole and its relative size with respect to the black hole is smaller for all values of $\gamma$ and $\beta$. For $\gamma=0$ the discs are infinite, irrespective of the value of $a$ and $\beta$. As $\gamma$ and $\beta$ increase from 0 to 0.9 (compare the three top rows) the discs become significantly thinner and radially smaller (see also Fig.~\ref{more-profile} below). It is also interesting to note that the maximum value of the rest-mass density is higher (with respect to the value at the disc centre) as the spin increases.

\begin{figure}[t]
\centering
\includegraphics[scale=0.2]{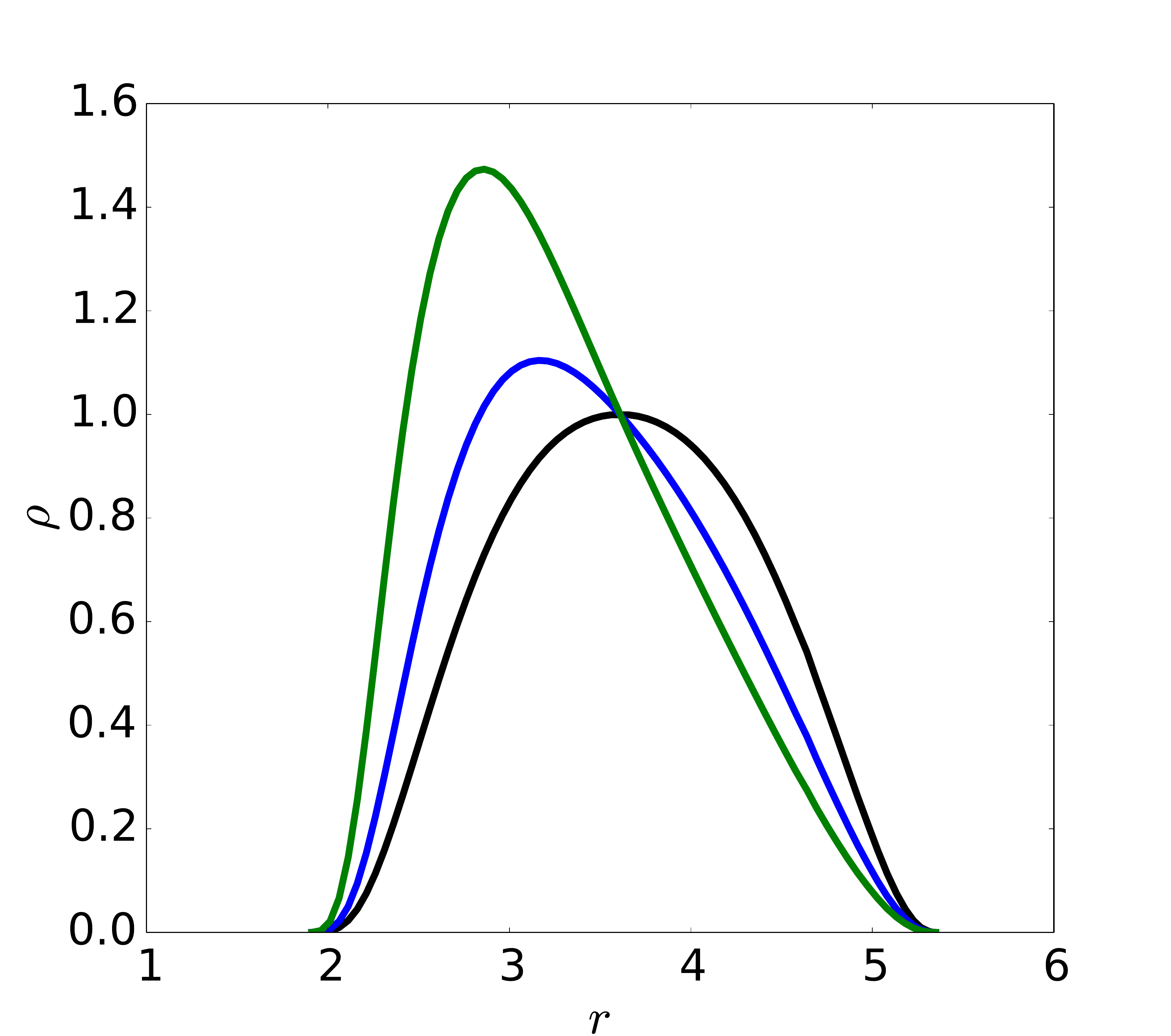}
\caption{Radial profiles of the rest-mass density in the equatorial plane for {$\beta_{\mathrm{m}_{\mathrm{c}}}=10^3$} (black curve), 1 (blue curve), and $10^{-3}$ (green curve). }
\label{magnetisation-profile}
 \end{figure}
 
\begin{figure}[t]
\centering
\includegraphics[scale=0.2]{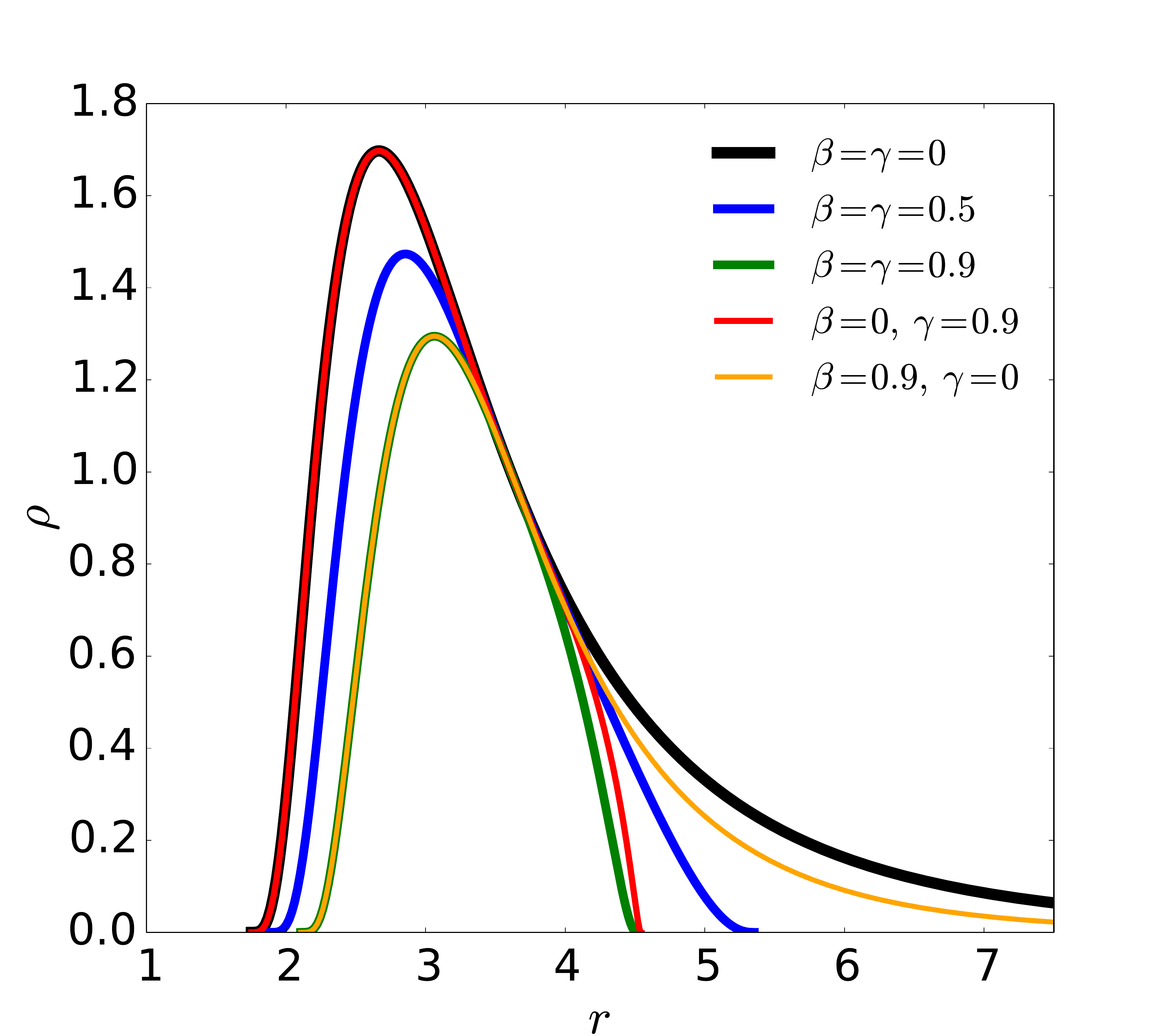}
\caption{Radial profiles at the equatorial plane for models with $a=0.9$, $\beta_{\mathrm{m}_{\mathrm{c}}}
=10^{-3}$, and same combination of the $\gamma$ and $\beta$ parameters as in the rows of Fig.~\ref{models} (the specific values are indicated in the legend).}
\label{more-profile}
\end{figure}

\begin{figure*}
\centering
\includegraphics[scale=0.14]{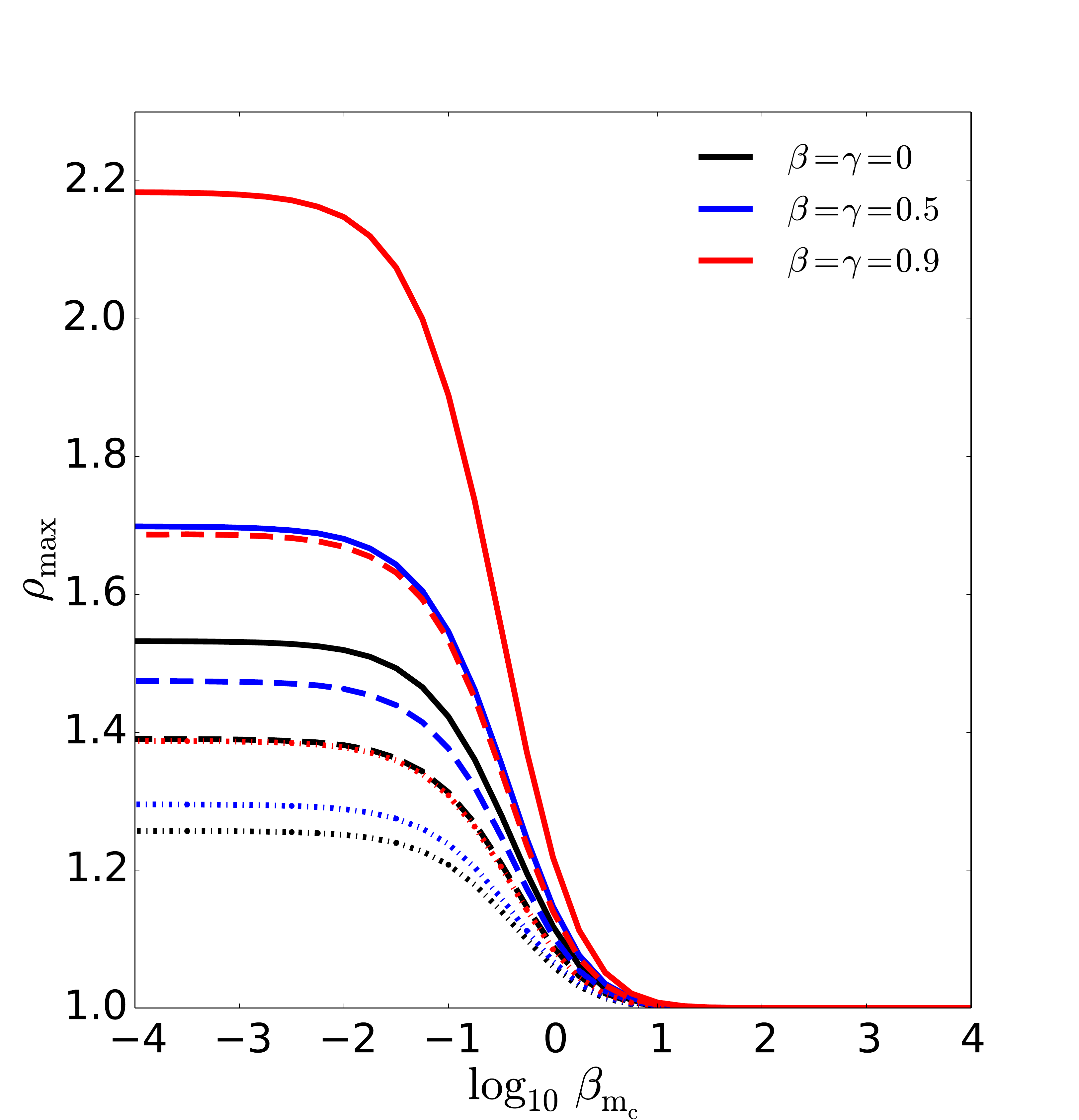}
\hspace{-0.3cm}
\includegraphics[scale=0.14]{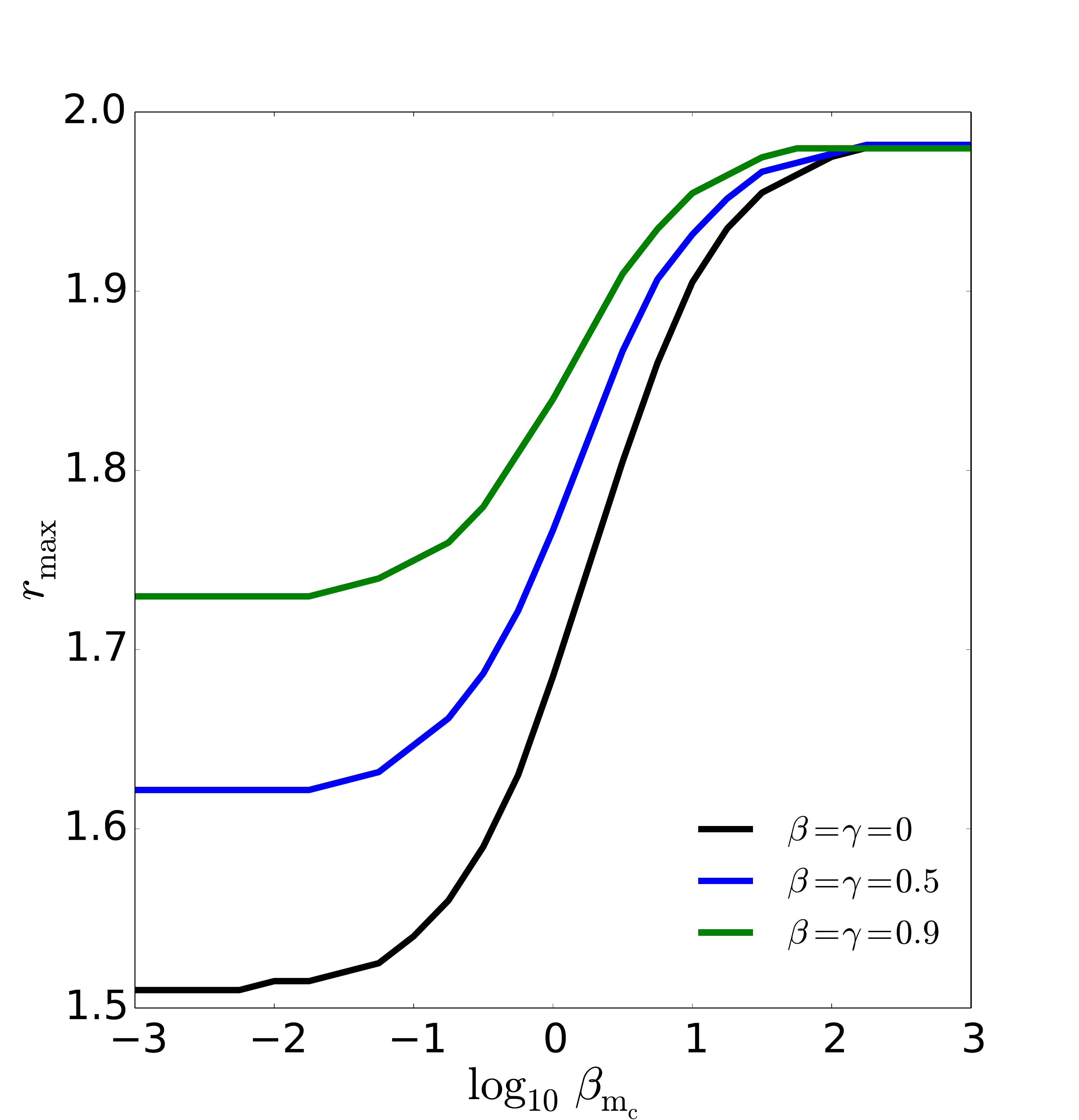}
\hspace{-0.2cm}
\includegraphics[scale=0.14]{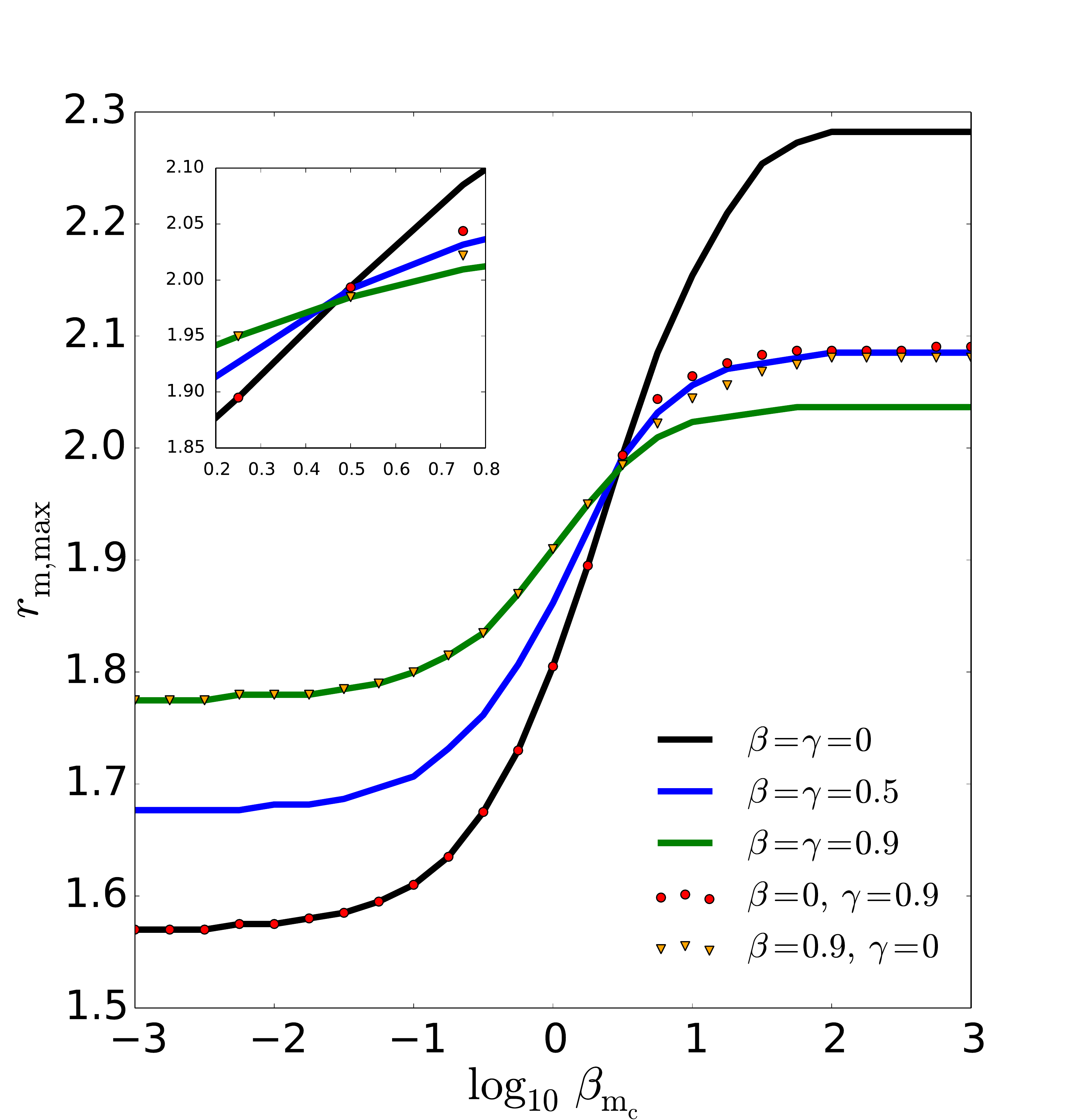}
\caption{Effects of $\beta_{\mathrm{m}_{\mathrm{c}}}$. Left panel: variation of the value of the maximum rest-mass density with respect to the logarithm of the magnetisation parameter. The solid, dashed, and dotted lines refer to $a = 0.99$, $0.9$ and $0.5$, respectively, and the colour code refers to the models shown in the legend. Middle panel: variation of the location of the rest-mass density maximum (and fluid pressure) with respect to the logarithm of the magnetisation parameter for the three models shown in the legend with $a = 0.99$. Right panel: variation (also for $a = 0.99$) of the location of the maximum of the magnetic pressure with respect to the logarithm of $\beta_{\mathrm{m}_{\mathrm{c}}}$.}
           \label{max-vs-magnetisation}%
 \end{figure*}
 
Figure~\ref{magnetisation} shows the effects of changing the parameter $\beta_{\mathrm{m}_{\mathrm{c}}}$ (the magnetisation) in the structure of the discs. From left to right the values plotted in each panel are $\beta_{\mathrm{m}_{\mathrm{c}}}=10^3$, 1, and $10^{-3}$. The model chosen corresponds to $a=0.9$ and $\gamma=\beta=0.5$. The larger the value of $\beta_{\mathrm{m}_{\mathrm{c}}}$ the less important the effects of the magnetisation in the disc structure. Figure~\ref{magnetisation} shows that, at least for the kind of magnetic field distribution we are considering, the effects of the magnetisation are minor. The disc structure remains fairly similar for all values of $\beta_{\mathrm{m}_{\mathrm{c}}}
$ and the only quantitative differences are found in the location of the centre of the disc (which moves inward with decreasing $\beta_{\mathrm{m}_{\mathrm{c}}}$) and of the range of variation of the isodensity contours (the maximum being slightly larger with decreasing $\beta_{\mathrm{m}_{\mathrm{c}}}$). This can be more clearly visualized in Figure~\ref{magnetisation-profile} which displays the radial profile at the equatorial plane of the rest-mass density for the same three cases plotted in Fig.~\ref{magnetisation}.

Next, we show in Fig.~\ref{more-profile} the corresponding radial profiles at the equatorial plane for the models with black hole spin $a=0.9$. We consider the case $\beta_{\mathrm{m}_{\mathrm{c}}}=10^{-3}$ (highest magnetisation) and the same combination of the $\gamma$ and $\beta$ parameters that we employed in Fig.~\ref{models}. Therefore, this plot depicts the radial profiles of the models occupying the central column of Fig.~\ref{models}. These profiles allow for a clearer quantification of the radial extent of the discs with $\gamma$ and $\beta$. As mentioned in the description of Fig.~\ref{models} as $\gamma$ and $\beta$ increase from 0 to 0.9 (compare black, blue, and green curves) the discs become gradually smaller. At the same time, the radial location of the rest-mass density maximum increases and the maximum value of the central rest-mass density (and pressure) decreases. It is worth noticing that the radial profiles of some of the models overlap below a certain radius. More precisely, the black and red curves overlap below $r\sim r_{\mathrm{c}}$ (which corresponds to $\rho \sim 1$) and also the green and orange curves. In addition, this figure clearly shows that, besides modifying the thickness of the disc, $\beta$ is the parameter which determines the value of the rest-mass density (and pressure) maximum. On the other hand, the parameter $\gamma$ is only relevant for controlling the radial size of the discs.

In order to provide a quantitative comparison of the structural differences that appear along the sequence of equilibrium models, we plot in Fig.~\ref{max-vs-magnetisation} the variation of the maximum of the rest-mass density (left panel), the radial location of the maximum of the fluid pressure (middle panel), and the radial location of the maximum of the magnetic pressure (right panel) as a function of the magnetisation parameter $\beta_{\mathrm{m}_{\mathrm{c}}}$. All quantities plotted are measured at the equatorial plane. Let us first consider the left panel. The line type used indicates the value of the black hole spin $a$, namely, a dotted line is for $a=0.5$, a dashed line for $a=0.9$, and a solid line for $a=0.99$. Correspondingly, the colour of the lines indicates the value of $\beta$ and $\gamma$ used, namely, red curves correspond to $\gamma=\beta=0$, the blue ones to $\gamma=\beta=0.5$, and the black ones to $\gamma=\beta=0.9$. All curves show the same monotonically decreasing trend with $\beta_{\mathrm{m}_{\mathrm{c}}}$, yet for small enough and large enough values of $\beta_{\mathrm{m}_{\mathrm{c}}} $ the value of the rest-mass density maximum does not change. However, in the interval $10^{-2}\le \beta_{\mathrm{m}_{\mathrm{c}}} \le 10^{2}$, $\rho_{\rm max}$ changes abruptly. The larger the black hole spin the larger the drop in the maximum of the rest-mass density. For sufficiently large values of $\beta_{\mathrm{m}_{\mathrm{c}}}$ (small magnetisation) the value of $\rho_{\rm max}$ stays constant to the same value irrespective of the black hole rotation.

In the middle panel of Fig.~\ref{max-vs-magnetisation} we show the radial location of the maximum of the fluid pressure as a function of $\beta_{\mathrm{m}_{\mathrm{c}}}$. The black hole spin is $a=0.99$ and the values of $\beta$ and $\gamma$ are indicated in the legend. For all values of $\beta$ and $\gamma$, the maximum of the location of the disc fluid pressure decreases with decreasing $\beta_{\mathrm{m}_{\mathrm{c}}}$. Below $\beta_{\mathrm{m}_{\mathrm{c}}}\sim 10^{-3}$ the location of the maximum does no longer change. The panel also shows that above $\beta_{\mathrm{m}_{\mathrm{c}}}\sim 10^{3}$ the constant value of $r_{\rm max}$ is the same irrespective of $\beta$ and $\gamma$, as expected, because for purely hydrodynamic discs, $r_{\mathrm{max}} = r_{\mathrm{c}}$.

The dependence of the location of the maximum of the magnetic pressure with $\beta_{\mathrm{m}_{\mathrm{c}}}$ is shown in the right panel of Fig.~\ref{max-vs-magnetisation}. While this location also decreases with decreasing $\beta_{\mathrm{m}_{\mathrm{c}}}$, the constant value achieved for values above $\beta_{\mathrm{m}_{\mathrm{c}}}\sim 10^{3}$ does depend on the specific values of $\beta$ and $\gamma$, contrary to what happens with the fluid pressure. It must be noted that the location of the maximum of the magnetic pressure is identical for all models considered when $\beta_{\mathrm{m}_{\mathrm{c}}} \equiv 1/(\lambda - 1) = 3$, as we show in Appendix~\ref{app_magmax}. At this value of $\beta_{\mathrm{m}_{\mathrm{c}}}$ all models cross at $r_{\mathrm{m}_{\mathrm{max}}} = r_{\mathrm{c}}$, as it is more clearly shown in the inset of the right panel of Fig.~\ref{max-vs-magnetisation}.

As a final remark we note that in the left and middle panels of Fig.~\ref{max-vs-magnetisation} we do not show the data for models with $\beta \neq \gamma$ as they coincide with the respective models with the same value of $\beta$. Moreover, the middle and right panels only show the data for $a=0.99$ because changing the value of the spin parameter of the black hole only yields a change of scale. Therefore, if we chose the range of the graph accordingly, the plots would be identical (same relative differences between the curves) irrespective of the value of $a$.

\section{Conclusions}
\label{conclusions}

In this paper we have presented a procedure to build equilibrium sequences of magnetised, non-self-gravitating discs around Kerr black holes which combines the two existing approaches of~\citet{Komissarov:2006} and~\citet{Qian:2009}. On the one hand we have followed~\citet{Qian:2009} and have assumed a form of the angular momentum distribution in the disc from which the location and morphology of surfaces of equal potential can be computed. As a limiting case, this ansatz includes the constant angular momentum case originally employed in the construction of thick tori -- or Polish doughnuts~\citep{Abramowicz:1978,Kozlowski:1978}  -- and was already used by~\citet{Qian:2009} to build equilibrium sequences of purely hydrodynamical models. On the other hand, our discs are endowed with a purely toroidal magnetic field, as in the work of~\citet{Komissarov:2006}, which provides the methodology we have followed to handle the magnetic terms. On a similar note, we cite the work of~\citet{Wielgus:2015} who have recently extended the solution of~\citet{Komissarov:2006} to include non-constant specific angular momentum tori. These authors limited their consideration to power-law distributions and were particularly focused on the stability of such tori to the MRI. The approach discussed in our work differs from that of~\citet{Wielgus:2015} and, moreover, we have explored a much wider parameter space.

We have discussed the properties of the new models and their dependence on the initial parameters, namely the magnetisation parameter $\beta_{\mathrm{m}_{\mathrm{c}}}$, the parameters $\beta$ and $\gamma$ describing the angular momentum distribution, the black hole spin parameter $a$, and the inner radius of the disc $r_{\mathrm{in}}$. We have shown the effects of changing $\beta$ and $\gamma$ beyond the purely hydrodynamical case considered in~\citet{Qian:2009}. The morphology of the solutions we have built no longer changes for magnetisation values above $\beta_{\mathrm{m}_{\mathrm{c}}} \sim 10^{3}$ and below $\beta_{\mathrm{m}_{\mathrm{c}}} \sim 10^{-3}$. These cases can thus be considered as the hydrodynamical and MHD limiting cases, respectively. The new sequences of magnetised discs around black holes presented in this work can be used as initial data for magnetohydrodynamical evolutions in general relativity. In the near future, we plan to extend this work along two main directions, namely (i) including the self-gravity of the discs, following the approach laid out in~\citet{Stergioulas:2011}, and (ii) constructing accretion discs around hairy black holes, both, with scalar and Proca hair~\citep{Herdeiro:2014,Herdeiro:2016}.

\begin{acknowledgements}
It is a pleasure to thank P.~C.~Fragile and N.~Stergioulas for useful discussions and comments. We thank the referee for his constructive report which improved the manuscript. Work supported by the Spanish MINECO (grant AYA2015-66899-C2-1-P) and the Generalitat Valenciana (PROMETEOII-2014-069).
\end{acknowledgements}

\bibliographystyle{aa}
\bibliography{references}

\begin{appendix}
\section{Divergence of equation~\eqref{eq:F} at $\theta = \pi/2$}\label{div_partial_W}
To prove that Eq.~\eqref{eq:F} diverges at $\theta = \pi/2$ we need to show that
Eq.~\eqref{eq:polar_der_pot} vanishes at the equator, i.e.
\begin{eqnarray}
\partial_{\theta} W = \partial_{\theta} \ln|u_t| - \frac{\Omega \partial_{\theta}l}{1 - \Omega l}\, = 0.
\end{eqnarray}
Then, we have to prove that either the two terms of this equation are equal or they are identically zero at $\theta = \pi/2$. We shall see that the latter is true.
First, we take the derivative of the angular momentum distribution $\partial_{\theta} l$. From Eq.~(7) we can write the angular momentum distribution as $l(r, \theta) = l(r) \sin^{2\gamma} \theta$, so that the derivative reads
\begin{eqnarray}
\partial_{\theta} l = 2\gamma l(r) \sin^{2\gamma - 1} \theta \cos \theta \,.
\end{eqnarray}
Taking $\theta = \pi/2$ leads to $\partial_{\theta} l = 0$, so the second term equals to zero.
To show that the first term is also zero, we write it as
\begin{eqnarray}
\partial_{\theta} \ln|u_t| = \partial_{\theta} \ln \left(\frac{\mathcal{L}}{\mathcal{A}}\right)^{\frac{1}{2}}\,,
\end{eqnarray}
where $\mathcal{L} = g_{t \phi}^2 - g_{tt} g_{\phi\phi}$ and $\mathcal{A} = g_{\phi\phi} + 2 l g_{t\phi} + l^2g_{tt}$, and we have dropped the absolute value, as it is irrelevant to this discussion. Then, the derivative is
\begin{eqnarray}
\partial_{\theta} \ln|u_t| = \frac{1}{2} \frac{\mathcal{A}}{\mathcal{L}}\frac{\mathcal{A}\partial_{\theta}{\mathcal{L}} - \mathcal{L}\partial_{\theta}\mathcal{A}}{\mathcal{A}^2} = \frac{1}{2} \frac{\mathcal{A}\partial_{\theta}{\mathcal{L}} - \mathcal{L}\partial_{\theta}\mathcal{A}}{\mathcal{A} \mathcal{L}}\,.
\end{eqnarray}
We use Boyer-Lindquist coordinates, for which the metric components read
\begin{eqnarray}
g_{tt} = - \left(1 - \frac{2Mr}{\rho^2}\right), \ g_{t\phi} = -\frac{2Mar\sin^2\theta}{\rho^2},
\nonumber \\
g_{\phi\phi} = \left(r^2 + a^2 + \frac{2Ma^2r\sin^2\theta}{\rho^2}\right) \sin^2 \theta\,,
\end{eqnarray}
where $\rho^2 = r^2 + a^2\cos^2 \theta$. Therefore, we obtain $\mathcal{L} = \Delta \sin^2 \theta$, where $\Delta = r^2 - 2Mr + a^2$, and its derivative $\partial_{\theta}\mathcal{L} = 2 \Delta \sin \theta \cos \theta$, which is zero for $\theta = \pi/2$. Next, we take the derivative of $\mathcal{A}$
\begin{eqnarray}
\partial_{\theta} \mathcal{A} = \partial_{\theta} g_{\phi\phi} + 2 (\partial_{\theta} l) g_{t\phi} + 2l\partial_{\theta} g_{t\phi} + 2l(\partial_{\theta} l) g_{tt} + l^2 \partial_{\theta} g_{tt} = 
\nonumber \\
= \partial_{\theta} g_{\phi\phi} + 2l\partial_{\theta} g_{t\phi} + l^2 \partial_{\theta} g_{tt}\,,
\end{eqnarray}
where we have used the previous result that $\partial_{\theta} l=0$ at $\theta = \pi/2$. By inspecting the metric components, it is easy to see that all terms depending on $\theta$ are functions of $\sin^2 \theta$ or $\cos^2 \theta$. This means their $\theta$ derivatives will have at least a $\cos \theta$ multiplying. Then $\partial_{\theta} \mathcal{A} = 0$ at $\theta = \pi/2$. Therefore, $\partial_{\theta} \ln|u_t|$ is also zero at $\theta = \pi/2$.

\section{Value of the magnetisation parameter for $r_{\mathrm{m}_{\mathrm{max}}} = r_{\mathrm{c}}$}\label{app_magmax}
In this appendix we derive the condition $\beta_{\mathrm{m}_{\mathrm{c}}} = 1/(\lambda - 1)$ for $r_{\mathrm{m}_{\mathrm{max}}} = r_{\mathrm{c}}$.
First, we can use Eq.~\eqref{eq:eos_mag} to write 
\begin{eqnarray}\label{eq_max}
\frac{\partial{p_{\mathrm{m}}(r)}}{\partial r} = \partial_{r} \left(K_{\mathrm{m}} \mathcal{L}^{\lambda -1} w^{\lambda}\right) = 0\,,
\end{eqnarray}
which yields
\begin{eqnarray}
\partial_{r} p_{\mathrm{m}} = K_{\mathrm{m}} \mathcal{L}^{\lambda - 2}w^{\lambda} [(\lambda - 1)(\partial_r \mathcal{L}) w + \lambda \mathcal{L} (\partial_r w)]=0\,.
\end{eqnarray}
Inside the disc $K_{\mathrm{m}} \mathcal{L}^{\lambda - 2}w^{\lambda} \neq 0$. Therefore, to fulfill the extremum condition~\eqref{eq_max} we need 
\begin{equation}\label{eq:simple_expr}
(\lambda - 1)(\partial_r \mathcal{L}) w + \lambda \mathcal{L} (\partial_r w) = 0\,.
\end{equation}
To evaluate this expression, we need to compute the partial derivatives $\partial_r \mathcal{L}$ and $\partial_r w$. The derivative of $\mathcal{L}$ is straightforward, since $\mathcal{L} = \Delta \sin^2 \theta$,
\begin{equation}
\partial_r \mathcal{L} = 2(r-M)\sin^2 \theta\,.
\end{equation}
Let us now discuss $\partial_r w$. From Eq.~\eqref{eq:enthalpy_eq} we have
\begin{equation}
w = \left[-\Delta W \left(\frac{\lambda -1}{\lambda}\right)\frac{1}{K+K_{\mathrm{m}}\mathcal{L}^{\lambda - 1}}\right]^{\frac{1}{\lambda-1}}\,.
\end{equation}
Taking its derivative we obtain
\begin{eqnarray}
\partial_r w &=& \frac{w}{\lambda - 1} \left[-\Delta W \left(\frac{\lambda -1}{\lambda}\right)\frac{1}{K+K_{\mathrm{m}}\mathcal{L}^{\lambda - 1}}\right]^{-1} 
\nonumber \\
&\times&
\left[\left(-\partial_r W \left(\frac{\lambda -1 }{\lambda}\right) \frac{1}{K+K_{\mathrm{m}}\mathcal{L}^{\lambda - 1}}\right) 
\right.
\nonumber \\
&+& \left. \left(-\Delta W \left(\frac{\lambda -1}{\lambda}\right)\frac{-(\lambda - 1) K_{\mathrm{m}} \mathcal{L}^{\lambda - 2} \partial_r \mathcal{L}}{(K+K_{\mathrm{m}}\mathcal{L}^{\lambda - 1})^2}\right)\right]\,.
\end{eqnarray}
Since at $r = r_{\mathrm{c}}$ the derivative of the potential is zero, $\partial_r W(r, \pi/2)|_{r = r_{\mathrm{c}}} = 0$, we can simplify the above expression to obtain
\begin{eqnarray}
\partial_r w &=& \frac{w}{\lambda - 1} \left[-\Delta W \left(\frac{\lambda -1}{\lambda}\right)\frac{1}{K+K_{\mathrm{m}}\mathcal{L}^{\lambda - 1}}\right]^{-1} 
\nonumber \\
&\times&
\left(-\Delta W \left(\frac{\lambda -1}{\lambda}\right)\frac{1}{K+K_{\mathrm{m}}\mathcal{L}^{\lambda - 1}}\right)
\nonumber \\
&\times&\left[\frac{-(\lambda - 1) K_{\mathrm{m}} \mathcal{L}^{\lambda - 2} \partial_r \mathcal{L}}{K+K_{\mathrm{m}}\mathcal{L}^{\lambda - 1}}\right]\,,
\end{eqnarray}
which can be further simplified to the following form
\begin{equation}
\partial_r w = - w \left[\frac{K_{\mathrm{m}} \mathcal{L}^{\lambda - 2} \partial_r \mathcal{L}}{K+K_{\mathrm{m}}\mathcal{L}^{\lambda - 1}}\right]\,.
\end{equation}
Next, by inserting this expression in Eq.~\eqref{eq:simple_expr} we obtain
\begin{equation}
(\partial_r \mathcal{L}) w [(\lambda -1)(K+K_{\mathrm{m}}\mathcal{L}^{\lambda - 1}) - \lambda K_{\mathrm{m}}\mathcal{L}^{\lambda - 1}] = 0\,.
\end{equation}
Since $(\partial_r \mathcal{L}) w \neq 0$ inside the disc, the above equation leads to 
\begin{equation}
K (\lambda - 1) = K_{\mathrm{m}}\mathcal{L}^{\lambda - 1}\,.
\end{equation}
Moreover, we can use Eqs.~\eqref{eq:eos_fluid}, \eqref{eq:eos_mag} and \eqref{eq:beta_eq} at the centre (remember that $w_{\mathrm{c}} = 1$) to write $K$ as $K = \beta_{\mathrm{m}_{\mathrm{c}}} K_\mathrm{m}\mathcal{L}^{\lambda -1}$. Therefore, this allows us to obtain the following simple expression for the value of the magnetisation parameter at the centre of the disc
\begin{equation}
\beta_{\mathrm{m}_{\mathrm{c}}} = \frac{1}{\lambda - 1}\,.
\end{equation}
It is relevant to note that the only reference to the explicit form of the metric is done to show that $\mathcal{L} \neq 0$ and $\partial_r \mathcal{L} \neq 0$ inside the disc. This means that our result holds for any stationary and axisymmetric metric if the aforementioned inequalities hold. Moreover, the result is also true for any angular momentum distribution that allows for the existence of a cusp and a centre (and $r_{\mathrm{c}} \neq r_{\mathrm{cusp}}$).

\end{appendix}

\end{document}